\documentclass[12pt]{article}

\usepackage{amsmath}
\usepackage{amssymb}
\usepackage{amsthm}
\usepackage{natbib}
\usepackage{multirow}
\usepackage[colorlinks,linkcolor=blue,urlcolor=blue,anchorcolor=blue,citecolor=blue]{hyperref}
\usepackage{color}
\usepackage[justification=justified,singlelinecheck=off]{caption}
\usepackage{array}
\usepackage{lineno,hyperref}
\usepackage{subfigure}
\usepackage{multirow}
\usepackage{bm}
\usepackage{color}
\usepackage{appendix}
\usepackage{graphicx}
\usepackage{subfigure}
\usepackage{epstopdf}

\usepackage{longtable}
\usepackage{booktabs}
\usepackage{mathrsfs}
\usepackage{float}
\usepackage{geometry}
\newcommand{\y}{\mathbf{y}}
\newcommand{\BC}{\mathcal{B}_\mathcal{C}}
\newcommand{\C}{\mathcal{C}}
\newcommand{\D}{\mathcal{D}}
\newcommand{\M}{\mathcal{M}}
\newcommand{\A}{\mathcal{A}}
\newcommand{\I}{\mathcal{I}}
\newcommand{\E}{\mathbb{E}}
\newcommand{\X}{\mathbf{X}}
\newcommand{\XC}{\mathbf{X}_\mathcal{C}}
\newcommand{\wXj}{\widetilde{X}_j}

\newcommand{\DA}{\mathcal{A}\cap\mathcal{D}}

\newcommand{\hM}{\widehat{\mathcal{M}}}

\newcommand{\mi}{\underset{i}{\max}}

\newcommand{\R}{\mathcal{R}}
\newcommand{\hR}{\widehat{\mathcal{R}}}

\newtheorem{theorem}{\sc Theorem}[section]
\newtheorem{lemma}{\sc Lemma}[section]
\newtheorem{proposition}{\sc Proposition}[section]

\newtheorem{remark}{\sc Remark}[section]

\title{Feature screening for multi-response linear models by empirical likelihood}

\author{Jun Lu$^1$, Qinqin Hu$^2$\footnote{The corresponding author: qqhu@sdu.edu.cn. The paper has been accepted by SCIENTIA SINICA Mathematica (in Chinese) }~~and Lu Lin$^{3}$
\\
\small{$^1$College of Liberal Arts and Sciences, National University of Defense Technology, Changsha, China}\\
\small{$^2$School of Mathematics and Statistics, Shandong University, Weihai, China}\\
\small{$^3$Zhongtai Securities Institute for Financial Studies, Shandong University, Jinan, China}\\
}

\date{}

\begin{document}
\maketitle
\begin{abstract}
This paper proposes a new feature screening method for the multi-response ultrahigh dimensional linear model by empirical likelihood. Through a multivariate moment condition, the empirical likelihood induced ranking statistics can exploit the joint effect among responses, and thus result in a much better performance than the methods considering responses individually. More importantly, by the use of empirical likelihood, the new method adapts to the heterogeneity in the conditional variance of random error. The sure screening property of the newly proposed method is proved with the model size controlled within a reasonable scale. Additionally, the new screening method is also extended to a conditional version so that it can recover the hidden predictors which are easily missed by the unconditional method. The corresponding theoretical properties are also provided. Finally, both numerical studies and real data analysis are provided to illustrate the effectiveness of the proposed methods.
\end{abstract}

\section{Introduction}

With the rapid development of science and technology, ultrahigh dimensional data with ``large $p$, small $n$'' are  frequently encountered in diverse fields, such as biomedical imaging, neuroscience, tomography, tumor classification, and so on. A central task for this kind of data is to lower the huge dimensionality to a moderate scale using fast and effective methods. With this purpose, \cite{fan2008sure} firstly proposed the concept of sure independence screening (SIS) to handle the ultrahigh dimensional data. Since then, a long list of relevant literatures has been proposed, which basically can be classified into two groups: the model-based and the model-free methods. For the former, a concrete model should be specified before the screening, typical literature include \cite{fan2008sure}, \cite{wang2009forward},  \cite{xu2014sparse} and \cite{wang2016high} for linear models, \cite{he2019robust} for elliptical copula regression model, \cite{fan2010sure}, \cite{fan2011nonparametric} and \cite{liu2014feature} for generalized linear models, additive models and varying coefficients models, respectively. The model-based methods usually enjoy high computational efficiency but have to bear the risk of model misspecification, which could lead to invalid screening results. To avoid such a risk, statisticians developed the model-free methods, some typical works include but are not limited to \cite{li2012feature}, \cite{cui2015modelfree}, \cite{lu2017model}, \cite{pan2019a} and the reference therein. Additionally, to reduce the negative effect caused by the complicated correlation among predictors, researchers also put forward some conditional screening methods, see  \cite{barut2015conditional}, \cite{hu2017conditional}, \cite{lin2016adaptive} and \cite{lu2017model}.

The concept of empirical likelihood (EL) is introduced by \cite{owen1988empirical} and further studied by \cite{qin1994empirical} and \cite{newey2004higher}. The empirical likelihood approach is very famous for its nice property of self-studentized, which means that it is a data-driven method without imposing strict distributional assumptions on variables. \cite{chang2013marginal} firstly introduced the empirical likelihood to SIS for the univariate response regression model, and proposed a new screening procedure named ELSIS by constructing an empirical likelihood ratio statistics as the new screening index. Due to the nice property of empirical likelihood, ELSIS is able to incorporate additionally the level of uncertainties associated with estimators and adapts to the heterogeneity in the conditional variance of random error. Motivated by the work of \cite{chang2013marginal}, \cite{hu2017conditional} generalized the ELSIS method to a conditional version. 

In this paper, we extend the idea of \cite{chang2013marginal} to the multi-response regression model. Ultrahigh dimensional data with multi-response are widespread in many applications, for example, in the analysis of the phenotype-genotype relationship, researchers always collect several genotypes simultaneously such as blood pressure, blood glucose, and some other body indices (seen as responses) and hundreds of thousands of Single Nucleotide Polymorphism (short for SNP, seen as the ultrahigh dimensional features). In such a study, the collected lots of responses are usually highly correlated or have a group structure. As a result, the multi-response linear model could be very useful to fit this kind of data.  Technically, to jointly consider the multiple responses and exploit the correlation information in them, we build the empirical likelihood function through a multivariate marginal moment condition and set the value of the empirical likelihood ratio function at zero as a filter. As a comparison, a natural extension of \cite{chang2013marginal} is to build the empirical likelihood function for each response individually and then aggregate them together as ranking statistics. Obviously, it can be expected that the newly proposed method taking the responses into consideration simultaneously would result in a more accurate screening result. 

It is also worth noting that there have been several papers considering the screening for the multi-response model. For example, the well-known DCSIS proposed by \cite{li2012feature} can be directly applied to the multi-response; \cite{li2017ultrahigh} proposed a projection-based screening method; \cite{lu2018feature} built a canonical correlation-based screening method for varying coefficient models. \cite{ma2020variable} developed a two-stage screening method for multi-response linear model; \cite{he2019on} proposed a rank canonical correlation-based screening procedure. Compared with these methods, our paper contributes to the following aspects. Firstly, we apply the empirical likelihood to the multi-response case, thus the new method is robust to the heteroscedasticity of the random error, and is able to exploit the correlation information among responses. Secondly, we extend the empirical likelihood induced method to a conditional version by the centralization technique, which can help to recover the hidden predictors. Besides, when the conditioning set is blind to us, we suggest a two-step screening approach to recruit the remaining active predictors. Finally, we prove that the sure screening property of the newly proposed screening method can be achieved with the model size controlled within a reasonable scale. Also, the corresponding theories for the conditional method are also provided.

The rest of the present paper is organized as follows. In Section 2, we give the details of the methodological development of the new screening procedure. Section 3 extends the method to a conditional version. Section 4 provides the theoretical properties of the method. Section 5 presents Monto Carlo simulations and a real data analysis. All proof of the main theoretical results is postponed to the Appendix.

\section{Screening method by empirical likelihood}
\subsection{Motivation to the new method}
\cite{chang2013marginal} introduced the empirical likelihood to the sure independence screening for the univariate response linear model called ELSIS, based on the moment condition $\E\{X_j(Y-X_j\beta_j^M)\}=0.$ Suppose a series of i.i.d samples $(\bm X_i,Y_i)_{i=1}^n$ from $(\bm X, Y)$, then the ranking statistics can be defined as
$$l_j(v)=2\sum_{i=1}^n\log\{1+\alpha g_{i,j}(v)\},$$
where $\alpha$ is the Lagrange multiplier, $g_{i,jk}=X_{ij}(y_{ik}-X_{ij}v)$ satisfying
$\sum_{i=1}^n\frac{g_{i,jk}(v)}{1+\alpha g_{i,jk}(v)}=0.$

To make ELSIS applicable in the multi-response case, one ways is to compute the empirical likelihood ratio (ELR) function for each response and aggregate them together. Let $l_{jk}(v)$ be ELR function of the $k$-th response, then we can generalize it to two forms by taking the average and the maximum of $l_{jk}(0)$ respectively over all responses, namely,
\begin{equation*}
	l^A_{j}(0)=\frac{1}{q}\sum_{k=1}^ql_{jk}(0)\mbox{~and~}l^M_{j}(0)=\max_{1\leq k\leq q}l_{jk}(0),
\end{equation*}
defined as ELSIS$_{avg}$ and  ELSIS$_{max}$, respectively.

Intuitively, ELSIS$_{avg}$ and  ELSIS$_{max}$ are inefficient because both them do not take the association among responses into consideration. We in the following conduct a numerical experiment to illustrate this point of view. Consider the following model
\begin{equation}\label{varied_q}
\left\{
\begin{split}
	Y_1&=X_1+\varepsilon_1,\\
	Y_2&=X_1+X_2+\varepsilon_2,\\
	&\cdots\\
Y_q&= X_1+X_2+\cdots +X_q + \varepsilon_q,
\end{split}\right.
\end{equation}
where $X_i\sim N(0,1)$ for $i=1,\cdots,p$ and $\varepsilon_j\sim N(0,1)$ for $j=1,\cdots,q$. By this design, the response is $q$-dimensional and $\{X_1,\cdots,X_q\}$ are active variables. We present the corresponding simulation results in Table \ref{multivariatey}. It can be seen with the growth of the dimension $q$ of the response, ELSIS$_{avg}$ and ELSIS$_{max}$ collapse very quickly, but MELSIS which will be developed next section performs very well. 
\begin{table}[!h]
	\centering
	\caption{The five representative quantiles ($5\%,25\%,50\%,75\%,95\%$) of minimal model size (MMS) under model (\ref{varied_q}) over 400 simulations with $(n,p)=(100,1000)$. Noting that MMS of ELSIS$_{com}$ means the minimal model size under which that all active predictors are recovered in any model at least once.}\label{multivariatey}
	\begin{tabular*}{\hsize}{@{}@{\extracolsep{\fill}}clrrrrr@{}}
		\toprule
		$q$ & Method  & $5\%$ & $25\%$ & $50\%$ & $75\%$ & $95\%$ \\ \midrule
		5   & MELSIS  &5.0&	5.0&	5.0&	5.0&	5.0\\
		& ELSIS$_{com}$ &5.0&	6.0&	9.0&	15.0&	36.3\\
		& ELSIS$_{avg}$ &6.0&	10.0&	20.0&	41.8&	96.2\\
		& ELSIS$_{max}$ &5.0&	5.0&	6.0&	10.3&	39.2\\
		10   & MELSIS  &10.0&	10.0&	10.0&	11.0&	15.1\\
		& ELSIS$_{com}$ &37.9&	72.5&	171.5&	283.3&	644.1\\
		& ELSIS$_{avg}$ &80.9&	148.3&	217.0&	315.5&	444.0\\
		& ELSIS$_{max}$ &23.0&	73.3&	116.5&	211.0&	529.0\\
		15   & MELSIS  &15.0&	15.8&	17.0&	20.0&	38.3\\
		& ELSIS$_{com}$ &160.1&	276.0&	443.0&	636.8&	889.2\\
		& ELSIS$_{avg}$ &193.9&	282.8&	388.0&	509.0&	687.3\\
		& ELSIS$_{max}$ &113.7&	204.5&	318.0&	473.5&	657.0\\
	\bottomrule
	\end{tabular*}	
\end{table}

\subsection{A new ranking index}
Consider the following linear model
\begin{equation}\label{mvlm}
\bf y = BX+\bm\varepsilon,
\end{equation}
where ${\bf y}=(Y_1,\cdots,Y_q)^\top\in\mathbb{R}^q$ is a $q$-dimensional response, $q$ is allowed to diverge to infinity at some certain rate, ${\bf B}=(\beta_{ij})_{1\leq i\leq q, 1\leq j\leq p}$ is a $q$ by $p$ coefficient matrix, and ${\bf X}=(X_1,\cdots,X_p)^\top\in\mathbb{R}^p$ is a $p$-dimensional predictor. Without loss of generality, we assume that each component in $\bf X$ has been standardized with zero expectation and unit variance.
Define the index set of active predictors (true model) as
\begin{equation*}
\mathcal{A} = \left\{1\leq j\leq p: \|\bm\beta_j\|_2\neq 0 \right\},
\end{equation*}
where $\bm\beta_j = (\beta_{1j},\cdots,\beta_{qj})^\top$ and $\|\cdot\|_2$ is the Eculidean norm. With the sparse condition, it is always assumed that $|\mathcal{A}|$ is much smaller than $p$, where $|\mathcal{A}|$ is the cardinality of $\mathcal{A}$. Correspondingly, we define $\mathcal{I}=\{1,\cdots,p\}\backslash \mathcal{A}$ as the index set of inactive predictors. We are intend to reduce the large model (\ref{mvlm}) to a moderate size such that $\mathcal{A}$ is included in it.

For the multivariate response, we write the marginal moment equation as
\begin{equation}\label{ee1}
\mathbb{E}\{X_j(\y-X_j\bm\beta_j^M)\}=\bm0_{q\times 1}
\end{equation}
for $\bm\beta_j^M=(\beta_{1j}^M,\cdots,\beta_{qj}^M)^\top$, $j=1,\cdots,p$, where $\bm0_{q\times 1}$ represents a $q$-dimensional vector with entries equal to zero. From now on, we suppress the subscript of $\mathbf{0}_{q\times 1}$ whenever there is no confusion. Under the independence rule, it has that $\bm\beta_j^M=\bm\beta_j$.

Let $({\bf X}_i, {\bf y}_i)_{i=1}^n$ be a set of i.i.d samples from $(\bf X, y)$, where ${\bf X}_i=(X_{i1},\cdots,X_{ip})^\top$ and ${\bf y}_i=(y_{i1},\cdots,y_{iq})^\top$. Based on (\ref{ee1}), the empirical likelihood can be established as
\begin{equation*}
EL_j(\bm v)=\sup\left\{\prod_{i=1}^nw_i:w_i\geq0,\sum_{i=1}^nw_i=1, \sum_{i=1}^nw_ig_{ij}(\bm v)=\bm0\right\}
\end{equation*}
for $j=1,\cdots,p$, where $g_{ij}(\bm v)=X_{ij}({\bf y}_i-X_{ij}\bm v)$. Consequently, the marginal empirical likelihood ratio can be defined as
\begin{equation*}
l_j(\bm v)=-2\log\{EL_j(\bm v)\}-2n\log n=2\sum_{i=1}^n\log\{1+\bm\alpha^\top g_{ij}(\bm v)\},
\end{equation*}
where $\bm\alpha$ is the Lagrange multiplier satisfying
\begin{equation*}
\bm0 = \sum_{i=1}^n\frac{g_{ij}(\bm v)}{1+\bm\alpha^\top g_{ij}(\bm v)}.
\end{equation*}
Similarly to \cite{chang2013marginal}, we take $l_j(\bm0)$ as a ranking index becase it will be small if $\bm\beta_j=\bm0$ but large otherwise.
Then for a predetermined threshold value $\gamma_{1n}$, the true model can be estimated as
\begin{equation*}
\widehat{\mathcal{A}}_{\gamma_{1n}}=\{1\leq j\leq p:l_j(\bm0)\geq \gamma_{1n}\}.
\end{equation*}
For simplicity, we name this new method as MELSIS, representing a multivariate extension of ELSIS.

\subsection{Comparison with the existing methods} \label{comparison}
It can be shown that $l_j(\bm0)$ exploits the correlation among responses while $l_j^A(0)$ and $l_j^M(0)$ cannot. For simplicity, let $V_{ik}=X_{ij}Y_{ik}$ and $\bm V_i=(V_{i1},\cdots,V_{iq})^\top$, then by Taylor expansion, $l_j(\bm0)$ can be expressed as 
\begin{equation}\label{taylor1}
	l_j(\bm0)=n\left[\frac{1}{n}\sum_{i=1}^n\bm V_i^\top\right]\left[\frac{1}{n}\sum_{i=1}^n\bm V_i\bm V_i^\top\right]^{-1}\left[\frac{1}{n}\sum_{i=1}^n\bm V_i\right]+o_p(1),
\end{equation}
while $l_j^A(0)$ and  $l_j^M(0)$ are respectively expressed as 
\begin{eqnarray}\label{taylor2}
	l_j^A(0)&=&n\left[\frac{1}{n}\sum_{i=1}^n\bm V_i^\top\right]\left[\mbox{diag}\left(\frac{1}{n}\sum_{i=1}^n\bm V_i\bm V_i^\top\right)\right]^{-1}\left[\frac{1}{n}\sum_{i=1}^n\bm V_i\right]+o_p(1),\mbox{~and}\label{taylor2}\\
	l_j^M(0)&=&\max\left\{\frac{\left(\sum_{i=1}^nV_{i1}\right)^2}{\sum_{i=1}^nV_{i1}^2},\cdots,\frac{\left(\sum_{i=1}^nV_{iq}\right)^2}{\sum_{i=1}^nV_{iq}^2}\right\}+o_p(1) \label{taylor3}
\end{eqnarray}
Comparing formula (\ref{taylor1}), (\ref{taylor2}) and (\ref{taylor3}), the main difference of $l_j^A(0)$ or $l_j^M(0)$ from $l_j(\bm0)$ is that the former two screening indices consider the response individually, thus the group structure among responses is neglected, while $l_j(\bm0)$ exploits the correlation among responses. As \cite{luo2020feature} claimed, the group structure in responses reflects the correlation among variables within the group, a statistic ignoring the correlation information is not sufficient. Of course, if the responses are mutually uncorrelated, $l_j(\bm0)$ will naturally degenerate to $l_j^A(0)$. Actually, from another perspective, $l_j(\bm 0)$ can be seen as the Mahalanobis distance between $\E X_j{\bf y}$ and $\bm0$ while $l_j^A(0)$ is the scalable Euclidean distance between them. It is well known that the Mahalanobis distance is more efficient than the Euclidean one.

Additionally, \cite{ma2020variable} and \cite{he2019on} also developed the screening procedure for the multi-response models, but both them implies a homogeneous assumption on the variance of the random error, otherwise the resulting screening results would be inefficient. The numerical results in Section 5 also confirms this viewpoint.

\section{Extension of MELSIS to CMELSIS}
\subsection{Motivation of CMELSIS}
{
In high dimensional data, sometimes the complicated correlation among active predictors may produce hidden variables, which means that some active predictors with large coefficients could have small marginal utilities because of the interaction effect among active predictors. For example, considering the following model,
\begin{equation}\label{case1}
\left\{
\begin{split}
	Y_1&=2X_1-2X_2+\varepsilon_1 \\
	Y_2&=4X_1+6X_2-9X_3+\varepsilon_2,
\end{split}
\right.
\end{equation}
where $X_j\sim N(0,1)$ for $j=1,\cdots,p$ with $\mathrm{Corr}(X_i,X_j)$ equal to $0.9+0.1I(i=j)$, and $\varepsilon\sim N(0,1)$. Simple calculation shows that $\E X_3\y=(0,0)^\top$, which implies that the empirical likelihood constructed based on $\E X_3\y$ will miss $X_3$. 

To address the problem, in this section, we extend the newly proposed method MELSIS to a conditional version, named CMELSIS. Before that, we provide a proposition below to motivate the CMELSIS. 
\begin{proposition}
If 
\begin{equation}\label{cor0}
	\frac{K\lambda_{\max}(\mathrm{cov}(\X_\A,\X_\I^\top)\mathrm{cov}(\X_\I,\X_\A^\top))}
{\lambda_{\min}(\mathrm{cov}(\X_\A,\X_\A^\top))}\leq \min_{j\in\A}\|\E X_j\y\|^2,
\end{equation}
where $K$ is the number of active predictors, $\lambda_{\max}({\bf M})\left(\lambda_{\min}({\bf M})\right)$ is the max(min) eigenvalue of $\bf M$,  then it has that  $\max_{j\in\I}\|\E X_j\y\|\leq \min_{j\in\A}\|\E X_j\y\|$.
\end{proposition}
The proof of the proposition is given in Supplement. This proposition indicates that to rank active predictors before the inactive ones, two conditions are required. One is that the numerator in the left side of (\ref{cor0}) should be small, which means that the correlation between $\X_\mathcal{A}$ and $\X_{\mathcal{I}}$ should not be strong. The other is that the denominator in the left side should be large, which implies that the correlation among the active predictors themselves should be small, otherwise, the hidden variable might arise. Still taking model (\ref{case1}) as an example, it has that $\lambda_{\min}(\mathrm{cov}(\X_\A,\X_\A^\top))=0.1$, this small value will make condition (\ref{cor0}) easily violated.

\subsection{Ranking index by centralization}

To recover the hidden predictors, the conditional screening approach is a commonly used method which has beed studied by several literatures, see for example, \cite{barut2015conditional} and \cite{yi2018model}. In our view, these method works because they can prevent the marginal utility of a hidden variable from being canceled out, via introducing the conditional variables.

Without loss of generality, assuming that the first $s_\C$ predictors in $\mathbf{X}$ are known in advance, i.e., $\mathbf{X}_{\mathcal{C}}=(X_1,\cdots,X_{s_\C})^\top$. When it is not available, we can set the first several predictors selected by MELSIS as the conditioning set. Correspondingly, denoting $\mathbf{X}_{\mathcal{D}}$ as the complement of $\mathbf{X}_{\mathcal{C}}$ in $\mathbf{X}$. Then, our goal is to recruit the active predictors in $\mathbf{X}_{\mathcal{D}}$, in other words, to identify the index set $\mathcal{A}\cap \mathcal{D}=\{j\in \mathcal{D}:\|\bm\beta_j\|_2\neq 0\}$. 
To this end, instead of using $\E X_j\y$ to construct the empirical likelihood, we revise the moment as $\E\widetilde{X}_j\y$, where $\widetilde{X}_j=X_j-\E(X_j|\mathcal{B}_{\mathcal{C},j}^\top\XC)$ is a centralized version of $X_j$, $\mathcal{B}_{\mathcal{C},j}$ is a matrix such that $X_j$ is independent of $\XC$ given $\mathcal{B}_{\mathcal{C},j}$. In the following, when there is no confusion, we neglect the dependence of $\mathcal{B}_{\mathcal{C},j}$ on the subscript $j$. The following proposition demonstrates the benefit of the centralization to some extent, although under a simple setting $\BC=\bm I$.

\begin{proposition} Under linear condition (\ref{LC}), if $\BC=\bm I$, then when 
\begin{equation}\label{cor1}
	\frac{K\lambda_{\max}\left(\mathrm{cov}(\widetilde{\X}_\A,\X_\I^\top)\mathrm{cov}(\X_\I,\widetilde{\X}_\A^\top)\right)}
{\lambda_{\min} (\mathrm{cov}(\widetilde{\X}_\A,\widetilde{\X}_\A))}\leq \min_{j\in\A}\|\E\widetilde{X}_k\y\|^2,
\end{equation}
where $\widetilde{\X}_\A=\X_\A-\E(\X_\A|\BC^\top\X_\C)$, it has that $\max_{j\in\I}\|\E\widetilde{X}_j\y\|\leq \min_{j\in\A}\|\E\widetilde{X}_j\y\|$.
\end{proposition}
The proof of this proposition is also given in Supplement. Regarding this proposition, we have the following remark.
\begin{remark}
The condition (\ref{cor1}) might be weaker than (\ref{cor0}), it allows the situation where the correlation among active predictors is large. Intuitively, when $\XC$ is highly correlated with $\X_\A$, $\widetilde{\X}_\A$ can be seen as the error term $\bf E$ in the model $\X_\A=f(\XC)+\bf E$, then by extracting the information of $\XC$, (\ref{cor1}) can be easier to be satisfied. For example, when $f(\XC)=\BC^\top\XC$ and $\bf E$ has a low correlation with $\X_\I$, then $\frac{\lambda_{\max}\left(\mathrm{cov}(\widetilde{\X}_\A,\X_\I^\top)\mathrm{cov}(\X_\I,\widetilde{\X}_\A^\top)\right)}{\lambda_{\min} (\mathrm{cov}(\widetilde{\X}_\A,\widetilde{\X}_\A^\top))}=\frac{\lambda_{\max}\left(\mathrm{cov}({\bf E},\X_\I^\top)\mathrm{cov}(\X_\I,{\bf E}^\top)\right)}{\lambda_{\min} (\mathrm{cov}({\bf E},{\bf E}^\top))}$ will be small. 
To have a direct insight into condition (\ref{cor0}) and (\ref{cor1}), we still take model (\ref{case1}) as illustration by setting $(n,p)=(100,500)$. We select the first $[n/\log n]=21$ variables ranked by MELSIS as $\X_\C$, simple calculation shows that the mean of $\frac{\lambda_{\max}(\mathrm{cov}(\X_\A,\X_\I^\top)\mathrm{cov}(\X_\I,\X_\A^\top))}{\lambda_{\min}(\mathrm{cov}(\X_\A,\X_\A^\top))}$ based on 200 simulations equals to 814.2 while the mean of  $\frac{\lambda_{\max}\left(\mathrm{cov}(\widetilde{\X}_\A,\X_\I^\top)\mathrm{cov}(\X_\I,\widetilde{\X}_\A^\top)\right)}{\lambda_{\min} (\mathrm{cov}(\widetilde{\X}_\A,\widetilde{\X}_\A))}$ equals to 2.8. It is also worth mentioning that if $\XC$ is independent of $\X_\A$, (\ref{cor1}) will degenerate to (\ref{cor0}), at this time, the centralization is unnecessary.
\end{remark}

Based on the moment $\E\widetilde{X}_j\y$, the conditional marginal empirical likelihood ratio function can be constructed as
\begin{equation*}
  l^c_j(\bm 0)=2\sum_{i=1}^n\log\{1+\bm\alpha^\top g^c_{ij}(\bm 0)\},
\end{equation*}
{where $g^c_{ij}(\bm0)=\widetilde{X}_{ij}\y_i$ with $\widetilde{X}_{ij}=X_{ij}-\E(X_{j}|\mathcal{B}_{\mathcal{C}}^\top\mathbf{X}_{i\mathcal{C}})$}, $\bm\alpha$ is the Lagrange multiplier satisfying
$
  \sum_{i=1}^n\frac{g^c_{ij}(\bm 0)}{1+\bm\alpha^\top g^c_{ij}(\bm 0)}=\bm0.
$

To construct the empirical likelihood in the sample level, we have to estimate the conditional expectation $\mathbb{E}(X_{j}|\mathcal{B}_{\mathcal{C}}^\top\mathbf{X}_{i\mathcal{C}}))$ in $g^c_{ij}(\bm0)$. Before that we need to determine $\BC$, to simplify and accelerate the whole screening procedure, we employ the sliced inverse regression (SIR, \cite{li1991sliced}) to estimate $\mathcal{B}_{\mathcal{C}}$. Simply speaking, SIR regresses the $\XC$ against $X_j$, then under the linearity condition (see below), it is proved that the centered regression curve $\E(\XC|X_j)$ is contained in the linear subspace spanned by $\BC\mbox{cov}(\XC)$. SIR is a very popular method in the sufficient dimension reduction field, one can refer to \cite{li1991sliced} for more details about SIR. Note that SIR needs the following linearity condition (LC),
	\begin{equation}\label{LC}
		\mathbb{E}(\X|\BC^\top\XC)=\XC^\top\BC
		{\mathrm{cov}(\BC^\top\XC)}^{-1}\mathrm{cov}(\BC^\top\XC, \X)
	\end{equation}
	for some matrix $\BC$. The linear condition is widely used in the dimension-reduction literature, for example Condition 3.1 in \cite{li1991sliced}, Condition (2.2) in \cite{wang2015estimation} and Condition (C2) in \cite{zhu2011model}. 
	
	Denote by $\widehat{\mathcal{B}}_\C$ the estimate of $\BC$ by SIR, it remains to estimate the conditional expectation $\widehat{\mathbb{E}}(X_j|\widehat{\mathcal{B}}_{\C}^\top\X_{i\C})$. One approach to estimating the conditional expectation is to use the nonparametric method, however, it will incur the problem of parameter selection and the possible curse of dimensionality. Instead, with the LC condition, we can simply estimate $\widehat{\mathbb{E}}(X_j|\widehat{\mathcal{B}}_{\C}^\top\X_{i\C})$ as 
	$
	\X_{i\C}^\top\widehat{\mathcal{B}}_\C \widehat{\mbox{cov}}
	\left(\widehat{\mathcal{B}}_\C^\top\XC\right)^{-1}\widehat{\mbox{cov}}(\widehat{\mathcal{B}}_\C^\top\XC, X_j).
	$
}

With the estimation $\widehat{g^c_{ij}}(\bm 0)=[X_{ij} - \widehat{\mathbb{E}}(X_j|\widehat{\mathcal{B}}_\C^\top\XC)]\mathbf{y}_i$, we use
$
  \widehat{l^c_j}(\bm 0)=2\sum_{i=1}^n\log\{1+\bm\alpha^\top \widehat{g^c_{ij}}(\bm 0)\}
$
as the ranking index. For a predetermined threshold value $\gamma_{2n}$, the true model is estimated as
\begin{equation*}
\widehat{\mathcal{A}\cap\mathcal{D}}_{\gamma_{2n}}=\{j\in\mathcal{D}:\widehat{l^c_j}(\bm0)\geq \gamma_{2n}\}.
\end{equation*}
For simplicity, we name the above method CMELSIS, representing a conditional extension of MELSIS.

\subsection{A two-step screening method: MRELS-CMRELS}\label{twostagesis}
The conditional screening procedure is able to overcome the screening problem caused by the complicated correlation among predictors but it depends on the selection of the conditioning set. If no prior information can be obtained, CMELSIS is inapplicable. In this circumstance, we can employ MELSIS first to pre-select some predictors as a conditional set and then perform CMELSIS to select the remaining active predictors. We name this method a two-step screening procedure.

From the two-step screening method, we can induce a \textit{sequential screening} method. Let $\mathbf{X}_{\mathcal{C}_k}$ be the predictors selected by the $k$-th screening procedure, then the next candidate predictor added into $\mathbf{X}_{\mathcal{C}_k}$, denoted as $\mathbf{X}_{\mathcal{C}_{k+1}}$, is the one which maximizes the empirical likelihood ratio $l^c_j(\bm 0, \mathbf{X}_{\mathcal{C}_k})$, where  $l^c_j(\bm v, \mathbf{X}_{\mathcal{C}_k})$ is the empirical likelihood ratio function established by setting $\mathbf{X}_{\mathcal{C}_k}$ as the conditioning set. We can repeat the above procedure until some stopping rule is achieved. The starting set $\mathbf{X}_{\mathcal{C}_1}$ can be set as the top ranked variable selected by MELSIS.

Practically, we would like to recommend practitioners use the two-step screening method. On the one hand, it is proved theoretically that CMELSIS has the sure screening property as long as the conditioning set contains some variables that are active or are strongly correlated with active ones. On the other hand, our numerical experience tells that the sequential method is unable to get a more accurate screening result than CMELSIS but incurs heavy computational burden.

\section{Theoretical properties}
In this section,  we study the theoretical properties of MELSIS and CMELSIS.
\subsection{Theoretical properties of MELSIS}
\label{section:SSP}
We first derive the sure screening properties of the proposed screening procedure and then give a bound on the size of the selected set of variables. Before that, we assume the following conditions.

\begin{itemize}
\item [(C1)] The random variables $Y_k\,(k=1,\ldots,q)$ have bounded variance. For any $j\in \A$, there exists $c_1>0$ and $\kappa\in[0,\frac{1}{2})$ such that
$
\|\mathbb{E}(X_j\y)\|_2\geq c_1n^{-\kappa}.
$

\item [(C2)] There are positive constants $K_1,K_2,\gamma_1$ and $\gamma_2$ such that
\begin{align*}
\mathbb{P}\{|X_j|>u\}\leq K_1 \exp \{-K_2u^{\gamma_1}\}\quad \text{and }\quad \mathbb{P}\{|Y_k|>u\}\leq K_1 \exp \{-K_2u^{\gamma_2}\},
\end{align*}
for $j=1,\ldots,p $ and $k=1,\ldots,q $ and any $u>0$.
\end{itemize}

Condition (C1) is an identification condition for the set $\A$, which is weaker than $|\mathbb{E}X_jY_k|\geq c_1n^{-\kappa}$ for $k=1,\cdots,q$. Similar condition is  assumed in \cite{chang2013marginal} and \cite{li2017ultrahigh}.
Condition (C2) ensures the large deviation results that are used to get the exponential convergence rate, it is also assumed in \cite{zhu2011model} and \cite{chang2013marginal}.

\begin{lemma}
Under Conditions (C1)-(C2), there exists a positive constant $C_1$  depending only on $K_1, K_2,\gamma_1$ and $\gamma_2$ given in Condition (C2) such that for any $j\in \A$ and $L \rightarrow \infty$,
{\begin{align*}
 \mathbb{P}\left\{l_j(\bm 0)<\frac{c_1^2n^{1-2\kappa}}{L^2}\right\} \leq \left\{\begin{array}{c}{\exp \left(-C_1 n^{1-2 \kappa-2r}\right)+\exp \left(-C_1 L^{\gamma}\right), \text { if }(1-2 \kappa-2r)(1+2\delta)<1} \\ {\exp \left(-C_1 n^{\frac{1-\kappa-r}{1+\delta}}\right)+\exp \left(-C_1 L^{\gamma}\right), \text { if }(1-2 \kappa-2r)(1+2\delta)\geq 1}\end{array}\right.
\end{align*}
where $\gamma=\frac{\gamma_1\gamma_2}{\gamma_1+\gamma_2}$ and  $\delta=\max\left\{\frac{2}{\gamma}-1,0\right\}$ and $r$ is the order of the dimension $q$ of response, that is, $q(n)=O(n^r)$ and satisfies the condition $0<r+\kappa<\frac{1}{2}$.
}
\label{lemma:lj0}
\end{lemma}

Lemma \ref{lemma:lj0} states that for all $j\in \A$, the diverging rate of $\ell_j(\bm 0)$ is not slower than $L^{-2}n^{1-2\kappa}$. If $j\notin \A$, according to the argument in \cite{owen2001empirical}, it can be shown that the corresponding $l_j(\bm 0)$ is $O_p(1)$. 
Let $L=n^{\frac{1}{2}-\kappa-\tau}$ for some $\tau\in (0,\frac{1}{2}-\kappa)$, then we obtain directly a more clear uniform result that the set $\A$ can be distinguished by examining the marginal empirical likelihood ratio $l_j(\bm 0)$ for $j=1,\ldots,p$. Moreover, we establish the sure screening property for our approach in the following theorem.

\begin{theorem}
Under Conditions (C1)-(C2), there exists a positive constant $C_1$  depending only on $K_1, K_2,\gamma_1$ and $\gamma_2$ given in Condition (C2) such that, for any $\tau\in \left(0,\frac{1}{2}-\kappa\right)$,
{
\begin{align*}
\max _{j \in \A} \mathbb{P}\left\{l_j(\bm 0)<c_{1}^{2} n^{2 \tau}\right\} \leq \left\{\begin{array}{l}{\exp \left\{-C_{1} n^{(1-2 \kappa-2r) \wedge \frac{(1-2 \kappa-2 \tau) \gamma}{2}}\right\}, \text { if }(1-2 \kappa-2r)(1+2\delta)<1} \\ {\exp \left\{-C_{1} n^{\frac{1-\kappa-r}{1+\delta} \wedge \frac{(1-2 \kappa-2 \tau) \gamma}{2}}\right\}, \text { if }(1-2 \kappa-2r)(1+2\delta)\geq 1}\end{array}\right.
\end{align*}
and, hence, if $\gamma_{1n}=c_{1}^{2} n^{2 \tau}$, we have that
\begin{align*}\mathbb{P}\left\{\A \subset \widehat{\A}_{\gamma_{1n}}\right\} \geq \left\{\begin{array}{l}{1-s\exp\left\{-C_{1}  n^{(1-2 \kappa-2r) \wedge \frac{(1-2 \kappa-2 \tau)\gamma}{2}}\right\},\quad i f(1-2 \kappa-2r)(1+2 \delta) < 1} \\ {1-s\exp\left\{-C_{1} n^{\frac{1-\kappa-r}{1+\delta} \wedge \frac{(1-2 \kappa-2 \tau) \gamma}{2}}\right\}, \quad i f(1-2 \kappa-2r)(1+2 \delta) \geq 1}\end{array}\right.
\end{align*}
where $\gamma=\frac{\gamma_1\gamma_2}{\gamma_1+\gamma_2}$ and
 $\delta=\max\left\{\frac{2}{\gamma}-1,0\right\}$, and $r$ is the order of the dimension $q$ of response, that is, $q(n)=O(n^r)$ and satisfies the condition $0<r+\kappa<\frac{1}{2}$, and $s=|\A|$, the size of the set of non-sparse elements.}
\label{theorem:lj0}
\end{theorem}

An efficient screening procedure does not only possess sure screening property, but also retains a small set of variables after thresholding. Theorem \ref{theorem:lj} below shows that our proposed procedure can control the size of the selected submodel very well. It also indicates that the dimension of response can diverge to infinity at a certain rate not larger than 
{$\frac{2}{7}\eta\wedge(\frac{1}{2}-\kappa)$.}

\begin{theorem}
Under Conditions (C1)-(C2), if  $\max_{j\notin A} \|\E(X_{j}{\bf y})\|_2=O_p(n^{-\eta})$ with $\eta>\kappa$, and  $\min_{j\notin A,1\leq k\leq q} |\E(X_{j}^2Y_{k}^2)|\geq c_2$ for some $c_2>0$, there exists a positive constant $C_2$ depending only on $K_1, K_2,\gamma_1$ and $\gamma_2$ given in Condition (C2) such that for any $j\notin \A$ and  {$\tau\in\left(\frac{1+r}{2}-\eta,\frac{1}{2}-\kappa\right)$ }and $\gamma_{1n}=c_{1}^{2} n^{2 \tau}$,
\begin{equation}\label{control_ms}
\mathbb{P}\left\{|\widehat{\A}_{\gamma_{1n}}|>s\right\}\leq p\exp\left(-C_2n^{B(\gamma,r,\eta,\breve{\eta})}\right),
\end{equation}
where $B(\gamma,r,\eta,\breve{\eta})$ is some constant determined by $\gamma,r,\eta,\breve{\eta}$, see Appendix for the details, $r$ is the order of the dimension $q$ of response, that is, $q(n)=O(n^r)$ and satisfies the condition {$0<r<\frac{2}{7}\eta\wedge(\frac{1}{2}-\kappa)$}  and  $\gamma=\frac{\gamma_1\gamma_2}{\gamma_1+\gamma_2}$,  $\breve{\eta}=\eta+\frac{2\tau}{3}-\frac{1}{3}$.
\label{theorem:lj}
\end{theorem}

\subsection{Theoretical properties of CMELSIS}
Theoretical properties of CMELSIS is studied in this part.
Compared with MELSIS, the proof of the sure screening property of CMELSIS is just replacing $X_{ij} \mathbf{y}_i$ in ${l_j}(\bm 0)$ with  $\widetilde{X}_{ij}\mathbf{y}_i$.
Similar to Condition (C1), the following condition is required:
\begin{itemize}
  \item [(C1')] The random variables $Y_k\,(k=1,\ldots,q)$ have bounded variance. For any $j\in \DA$, there exists $c_3>0$ and $\kappa\in[0,\frac{1}{2})$ such that
\begin{align*}
\|\mathbb{E}\{X_j-\E(X_j|\BC^\top\X_\C)\}{\bf y}\|_2\geq c_3n^{-\kappa}.
\end{align*}
\end{itemize}

The following lemma shows the goal set $\DA$ can be clearly distinguished by ${l^c_j}(\bm 0)$.
\begin{lemma}
Under Conditions (C1') and (C2), there exists a positive constant $C_3$  depends only on $K_1, K_2,\gamma_1$ and $\gamma_2$ given in Condition (C2) such that, for any $\tau\in (0,\frac{1}{2}-\kappa)$,
{
\begin{align*}
\max _{j \in  \D\cap\A} \mathbb{P}\left\{l_j^c(\bm 0)<c_{3}^{2} n^{2 \tau}\right\} \leq \left\{\begin{array}{c}{\exp \left\{-C_{3} n^{(1-2 \kappa-2r) \wedge \frac{(1-2 \kappa-2 \tau) \gamma}{2}}\right\}, \text { if }(1-2 \kappa-2r)(1+2\delta)<1} \\ {\exp \left\{-C_{3} n^{\frac{1-\kappa-r}{1+\delta} \wedge \frac{(1-2 \kappa-2 \tau) \gamma}{2}}\right\}, \text { if }(1-2 \kappa-2r)(1+2\delta)\geq 1}\end{array}\right.
\end{align*}
where $\gamma=\frac{\gamma_1\gamma_2}{\gamma_1+\gamma_2}$ and
 $\delta=\max\left\{\frac{2}{\gamma}-1,0\right\}$, $r$ is the order of the dimension $q$ of response, that is, $q(n)=O(n^r)$ and satisfies the condition $0<r+\kappa<\frac{1}{2}$.
 }
\label{lemma:ljc}
\end{lemma}
Following Lemma~\ref{lemma:ljc}, the theorem below proves the sure screening property of CMELSIS.
\begin{theorem}
Under Conditions (C1') and (C2), if $\mi |X_{ij}Y_{ik}|=O_p(n^\omega)$ where $\omega<{1/2}-\kappa$, $j\in\C$ and $1\leq k\leq q$, there exists a positive constant $C_4$  depends only on $K_1, K_2,\gamma_1$ and $\gamma_2$ given in Condition (C2) such that, for any $\tau\in \left(0,\frac{1}{2}-\kappa\right)$,
{
\begin{align*}
\max _{j \in \D\cap\A} \mathbb{P}\left\{\widehat{l_j^c}(\bm 0)<c_{3}^{2} n^{2 \tau}\right\} \leq \left\{\begin{array}{c}{\exp \left\{-C_{4} n^{(1-2 \kappa-2r) \wedge \frac{(1-2 \kappa-2 \tau) \gamma}{2}}\right\}, \text { if }(1-2 \kappa-2r)(1+2\delta)<1} \\ {\exp \left\{-C_{4} n^{\frac{1-\kappa-r}{1+\delta} \wedge \frac{(1-2 \kappa-2 \tau) \gamma}{2}}\right\}, \text { if }(1-2 \kappa-2r)(1+2\delta)\geq 1}\end{array}\right.
\end{align*}
and, hence, if $\gamma_{2n}=c_{3}^{2} n^{2 \tau}$, we have that
\begin{align*}\mathbb{P}\left\{\DA\subset \widehat{\DA}_{\gamma_{2n}}\right\} \geq \left\{\begin{array}{l}{1-s_{\DA}\exp\left\{-C_{4}  n^{(1-2 \kappa-2r) \wedge \frac{(1-2 \kappa-2 \tau)\gamma}{2}}\right\},\quad i f(1-2 \kappa-2r)(1+2 \delta) < 1} \\ {1-s_{\DA}\exp\left\{-C_{4} n^{\frac{1-\kappa-r}{1+\delta} \wedge \frac{(1-2 \kappa-2 \tau) \gamma}{2}}\right\}, \quad i f(1-2 \kappa-2r)(1+2 \delta) \geq 1}\end{array}\right.
\end{align*}
where $\gamma=\frac{\gamma_1\gamma_2}{\gamma_1+\gamma_2}$ and
$\delta=\max\left\{\frac{2}{\gamma}-1,0\right\}$, $r$ is the order of the dimension $q$ of response, that is, $q(n)=O(n^r)$ and satisfies the condition $0<r+\kappa<\frac{1}{2}$, and $s_{\DA}=|\DA|$ is size of the set of non-sparse elements. 
}
 \label{theorem:lj0hat}
\end{theorem}

The following theorem shows that CMELSIS takes good control of the model size.
\begin{theorem}
	Under Conditions (C1') and (C2), if $\underset{{j\notin \DA}}\max\|\mathbb{E}\{[X_j-\mathbb{E}(X_j|\BC^\top\X_\C)]{\bf y}\}\|_2=O(n^{-\eta})$ where $\eta >\kappa$ and
	$\underset{j\notin \DA,1\leq k\leq q}\min\mathbb{E}\{[X_j-\mathbb{E}(X_j|\BC^\top\X_\C)]^2Y_k^2\}\geq c_4 $ for some $c_4>0$, there exists a positive constant $C_5$ such that, for any $j\notin \DA$ and any $\tau \in (\frac{1+r}{2}-\eta,
	\frac{1}{2}-\kappa)$ and $\gamma_{2n}=c_{3}^{2} n^{2 \tau}$,      
	\begin{equation*}
	\mathbb{P}\left\{|\widehat{\DA}_{\gamma_{2n}}|>s_{\DA}\right\}\leq p_1\exp\left(-C_5n^{B(\gamma,r,\eta,\breve{\eta})}\right),
	\end{equation*}
	where $B(\gamma,r,\eta,\breve{\eta})$ is the same constant in Theorem 4.2,  $p_1$ is the size of $X_\D$, and $\breve{\eta}=\eta+\frac{2\tau}{3}-\frac{1}{3}$,  and $\omega$ satisfies $\mi |X_{ij}Y_{ik}|=O_p(n^\omega)$ and $\omega<\frac12-\eta$, for $j\in\C$ and $1\leq k\leq q$ and  $r$ satisfies the dimension $q$ of response satisfies $q(n)=O(n^r)$ and $0\leq r<\frac{2}{7}\eta\wedge\frac{1}{2}-\kappa\wedge(\frac{1}{2}-2\omega)$, $C_5$ depends only on $K_1, K_2,\gamma_1$ and $\gamma_2$ given in Condition (C2).
	\label{theorem:lj0DAbar}
\end{theorem}

\section{Numerical studies}
\subsection{General simulation settings}
In this section, we conduct several numerical experiments to evaluate the performance of the proposed methods in various model settings.
We first check the effectiveness of MELSIS. The competitors include PS \citep{li2017ultrahigh}, ELSIS$_{avg}$, ELSIS$_{max}$, DCSIS, BCorSIS \citep{pan2019a} and RCC \citep{he2021on}(RCC$_{sp}$ represents the Spearman correlation induced CC and RCC$_{kd}$ represents the Kendall's $\tau$ correlation induced CC).
Sequentially, we examine the validity of CMELSIS with a model having hidden variables. We also compare the CMELSIS with CELSIS$_{avg}$, CELSIS$_{max}$, DCSIS and BCorSIS \citep{pan2019a}, { where CELSIS$_{avg}$ and CELSIS$_{max}$ are the conditional versions of ELSIS$_{avg}$ and ELSIS$_{max}$.}

We repeat each experiment 400 times and employ the following criteria to evaluate the performance of each method: 
(1) MMS, the minimum model size (MS) of the selected model that includes all the active predictors. We report the $5\%, 25\%, 50\%, 75\%$ and $95\%$ of MMS over 400 replications; 
(2) $P_j$, the percentage of submodel $\widehat{\mathcal{A}}$  with size $d_n$ that contains $X_j$ across 400 repeated experiments; 
(3) $P_a$, the percentage of submodel $\widehat{\mathcal{A}}$  with size $d_n$ that contains all true predictors across 400 repeated experiments.

{\it Selection of the thresholding value.} 
Roughly speaking, we can take the hard thresholding rule and soft thresholding rule to determine the model size. For the former, we can set $d_n=c[n/\log(n)]$ such as \cite{fan2008sure}, where $[a]$ means the integer part of $a$. For the latter, a commonly used approach is to  randomly switch the rows of $\y$ and re-compute the marginal utilities of all predictors as the auxiliary statistics, and set the upper $\tau$-quantiles of them as the thresholding value, for example, the $100\%$-quantile (maximum) or $99\%$-quantiles of the auxiliary marginal utilities.
The hard rule is easy to implement and can control the model size precisely but the choice of $d_n$ is not explainable, contrarily, the soft rule seems more reasonable but usually spends more computational cost and results in random model size. 

 Throughout the simulation, to compare the results $P_j$ of different methods under certain model size, we always select the hard thresholding rule and set $d_n=[n/\log n]$ unless otherwise specified. Also, for practical consideration, we examine the performance of different methods under the soft thresholding rule using Example 4.1 and Example 4.3. During the implementation of our screening methods, we always use the adjusted empirical likelihood (AEL) algorithm proposed by \cite{chen2008adjusted} to solve the optimization problem. 
 {To get $\BC$, we employ the R package {\it dr} to conduct the SIR procedure, in which the number of slices is selected as 9, which is the default settings in {\it dr}, and the number of directions is set as $b$ such that the sum of first $b$ eigenvalues of the weighted sample covariance matrix accounts for more than $80\%$ of the sum of all eigenvalues. According to our experience, in most situations, the threshold is achieved by only selecting the first two eigenvalues. }

\subsection{Monto Carlo simulations}
\textbf{Example 4.1.}
This example uses a simple linear model to examine the effectiveness of MELSIS. The model is set up as
\begin{equation*}
	\left\{\begin{split}	
		&Y_1 = 3X_1+2X_2+\varepsilon_1;\\
		&Y_2 = 4X_1+X_3+\varepsilon_2;\\
		&Y_3 = 2X_2+4X_4+\varepsilon_3;\\
		&Y_4 = 3X_4+X_5+\varepsilon_4,
	\end{split}\right.
\end{equation*}
where $X_j$ for $j=1,\cdots,p$ are independently from $N(0,1)$,  $\varepsilon_i=\sigma_i(\mathbf{X})\epsilon_i$ for $i=1,\cdots,4$ are the random errors with $\epsilon_i\sim N(0,1)$ and $\mathrm{corr}(\epsilon_i,\epsilon_j)=\rho$. We set $\rho$ equal to 0 and 0.5 representing independent random errors and correlated random errors, respectively. 
For $\sigma_i(\mathbf{X})$, we consider two types of errors:
(a) $\sigma_i(\mathbf{X})=1\mathrm{~for~} i=1,\cdots,4$;
(b) $\sigma_1(\mathbf{X})=1/(X_1+X_2), \sigma_3(\mathbf{X})=1/(X_2^2+X_4^2),\sigma_2(\mathbf{X})=\sigma_4(\mathbf{X})=1$. Thus the error variance is homogeneous in (a) and heterogeneous in (b).
By this design, we can see that the responses are tied together through some common predictors. $X_3$ and $X_5$ can be seen as two variables with weak signals because they have relatively small coefficients and appear in the model only once.
Table \ref{proportion_ex1}-\ref{ms_ex1} report the simulation results under the setting $(n,p)=(100,2000)$. We see that only PS and MELSIS can achieve the sure screening while the other methods behave poorly. More specifically, the following conclusions can be summarized from the tables:
\begin{enumerate}
	\item [(1)] Table \ref{proportion_ex1} shows that under the hard-thresholding rule, DCSIS, BCorSIS, ELSIS$_{avg}$ and ELSIS$_{max}$ cannot select $X_3$ and $X_5$ into model, these methods have poor performance in terms of $P_a$. Note that some of them are the model-free methods. The above phenomenon implies that jointly considering the responses can significantly improve the screening results.
	\item [(2)] When the error variance is homogeneous, PS and RCC has a slight better performance than MELSIS, however, in the case of heterogenous situation, the conclusion is reversed, Table \ref{ms_ex1} shows that under this situation the MMS of our method is much smaller than that of PS. Overall, the heteroscedasticity has a negative effect on all methods but our method suffers the least.
	\item [(3)] Table \ref{proportion_ex13_soft} also displays the numerical results under the soft thresholding rule, from which we can see that for the unconditional screening methods, the soft thresholding rule also results in a good performance, but the heteroscedasticity brings some negative effects on the new method. It can be seen that the model size determined by the soft-thresholding rule is much larger than $[n/\log n]$.
	\item [(4)]  It seems that correlated random errors  does not have an obvious impact on all methods.
\end{enumerate}

\textbf{Example 4.2.} This example employs a more general model to investigate the effectiveness of MELSIS. The model is formulated as follows:
\begin{equation*}	
Y_i = \sum_{i=1}^p\beta_{ij}X_j + \varepsilon_i \mathrm{~for~} i=1,2,\cdots,5,\\
\end{equation*}
where $\beta_{ij}=0$ for $j>5$ and $\beta_{ij}=UW$ for $j=1,\cdots,5$ with $U$ taking values $\pm 1$ and $0$ with probability $0.4$ and $0.2$, respectively,  and $W\sim Uniform(0,1)$, the predictor $X_j$ follows the standard normal distribution with $\mathrm{corr}(X_j, X_{j'})=0.3$ for $j\neq j'$. The random errors are generated in a similar way as Example 4.1 with (a) a homoscedastic random error  $\sigma_i(\mathbf{X}) =1\mathrm{~for~} i=1,\cdots,5$ and (b) heteroscedastic random error $\sigma_i(\mathbf{X})=1/X_i$ for $i=1,3,5$ and $\sigma_i(\mathbf{X}) =1\mathrm{~for~} i=2,4$.
Different from the model settings in Example 4.1 where some active predictors are weak signals, the important features in this model almost contribute equally to the response.
The simulation results are still presented in Table \ref{ms_ex2} under the setting $(n, p)= (200,1000)$.
From the tables, in addition to some similar conclusions to Example 4.1 can be observed, we have another two findings. First of all, the heteroscedasticity in random error has obvious negative effect on all screening methods except ours, it can be seen that DCSIS fails to recover the active predictors, BCorSIS misses $X_2$ and $X_4$, and PS misses $X_5$. Second, unlike the phenomenon observed in the previous example, here ELSIS$_{avg}$ has a superior performance than ELSIS$_{max}$, it is reasonable because the active predictors almost contribute equally to the response.

\begin{table}[!h]
	\centering
	\caption{The proportion of active predictors being selected in Example 4.1, under the hard thresholding rule.}\label{proportion_ex1}
	\begin{tabular}{lcccccccccccc}
	\toprule
	&\multicolumn{6}{c}{$\sigma_i(\mathbf{X})$:~case (a)}&\multicolumn{6}{c}{$\sigma_i(\mathbf{X})$:~case (b)}\\
	\cmidrule(lr){2-7}\cmidrule(lr){8-13}
	Method &$P_1$&$P_2$&$P_3$&$P_4$&$P_5$&$P_a$&  $P_1$&$P_2$&$P_3$&$P_4$&$P_5$&$P_a$ \\  \midrule
	\multicolumn{13}{l}{$\rho=0$}\\
	DCSIS        & 1.00 & 0.98 & 0.11 & 1.00 & 0.05 & 0.00 & 0.68 & 0.34 & 0.02 & 0.52 & 0.01 & 0.00 \\
	BCorSIS      & 1.00 & 0.91 & 0.06 & 1.00 & 0.03 & 0.00 & 0.01 & 0.02 & 0.03 & 0.02 & 0.01 & 0.00 \\
	PS           & 1.00 & 1.00 & 0.98 & 1.00 & 1.00 & 0.97 & 1.00 & 0.59 & 1.00 & 0.63 & 1.00 & 0.44 \\
    RCC$_{sp}$   & 1.00 & 1.00 & 0.90 & 1.00 & 0.98 & 0.96 & 1.00 & 1.00 & 0.85 & 1.00 & 0.90 & 0.79\\
    RCC$_{kd}$   & 1.00 & 1.00 & 0.56 & 1.00 & 0.91 & 0.52 & 1.00 & 1.00 & 0.55 & 1.00 & 0.85 & 0.50 \\
	ELSIS$_{avg}$& 1.00 & 0.99 & 0.08 & 1.00 & 0.15 & 0.02 & 1.00 & 0.99 & 0.10 & 0.97 & 0.20 & 0.02 \\
	ELSIS$_{max}$& 1.00 & 1.00 & 0.24 & 1.00 & 0.47 & 0.12 & 1.00 & 0.99 & 0.18 & 0.98 & 0.39 & 0.08 \\
	MELSIS       & 1.00 & 1.00 & 0.94 & 1.00 & 1.00 & 0.94 & 1.00 & 0.96 & 1.00 & 0.93 & 1.00 & 0.88 \\
	\multicolumn{13}{l}{$\rho=0.5$}\\
	DCSIS        & 1.00 & 1.00 & 0.08 & 1.00 & 0.08 & 0.01 & 0.68 & 0.36 & 0.03 & 0.57 & 0.04 & 0.00 \\
	BCorSIS      & 1.00 & 0.92 & 0.06 & 1.00 & 0.06 & 0.00 & 0.02 & 0.02 & 0.01 & 0.01 & 0.03 & 0.00 \\
	PS           & 1.00 & 1.00 & 1.00 & 1.00 & 1.00 & 1.00 & 1.00 & 0.63 & 1.00 & 0.68 & 1.00 & 0.49 \\
    RCC$_{sp}$   & 1.00 & 1.00 & 0.98 & 1.00 & 1.00 & 0.98 & 1.00 & 0.93 & 0.83 & 1.00 & 1.00 & 0.77 \\
    RCC$_{kd}$   & 1.00 & 1.00 & 0.67 & 1.00 & 0.96 & 0.65 & 1.00 & 0.91 & 0.59 & 1.00 & 0.98 & 0.56 \\
    ELSIS$_{avg}$& 1.00 & 1.00 & 0.06 & 1.00 & 0.17 & 0.01 & 1.00 & 0.98 & 0.07 & 0.98 & 0.23 & 0.03 \\
	ELSIS$_{max}$& 1.00 & 1.00 & 0.21 & 1.00 & 0.48 & 0.11 & 1.00 & 0.98 & 0.14 & 0.98 & 0.42 & 0.07 \\
	MELSIS       & 1.00 & 1.00 & 0.98 & 1.00 & 1.00 & 0.98 & 1.00 & 0.95 & 1.00 & 0.94 & 1.00 & 0.89 \\\bottomrule
	\end{tabular}
	\end{table}

	\begin{table}[!h]
		\centering
		\caption{The quartiles of minimum model size of the selected models in Example 4.1.}\label{ms_ex1}
		\scalebox{0.95}{
		\begin{tabular}{lrrrrrrrrrr}
			\toprule
			&\multicolumn{5}{c}{$\sigma_i(\mathbf{X})$:~case (a)}&\multicolumn{5}{c}{$\sigma_i(\mathbf{X})$:~case (b)}\\
			\cmidrule(lr){2-6}\cmidrule(lr){7-11}
			Method &$5\%$&$25\%$&$50\%$&$75\%$&$95\%$&  $5\%$&$25\%$&$50\%$&$75\%$&$95\%$ \\  \midrule
			\multicolumn{11}{l}{$\rho=0$}\\
			DCSIS        &90.0 &247.3  &462.5  &777.3  &1339.3  &220.6 &735.5  &1196.5 &1610.8 &1927.0  \\
			BCorSIS      &266.3&726.5  &1057.0 &1522.0 &1895.5  &939.5 &1540.5 &1791.5 &1894.5 &1977.3 \\
			PS           &5.0  &5.0    &5.0    &5.0    &11.0    &5.0   &5.0    &45.0   &426.3  &1573.8 \\
            RCC$_{sp}$   &5.0  &5.0    &5.0    &7.0    &95.1    &5.0   &5.0    &6.0    &12.0   &131.2\\
            RCC$_{kd}$   &6.0  &10.0   &21.5   &81.3   &317.7   &5.0   &9.0    &22.5   &104.5  &330.2\\
			ELSIS$_{avg}$&63.0 &175.0  &353.0  &542.25 &953.5   &44.0  &193.0  &375.0  &618.5  &1130.5 \\
			ELSIS$_{max}$&11.0 &72.5   &195.0  &405.5  &982.9   &17.0  &116.0  &310.5  &627.8  &1190.1\\
			MELSIS       &5.0  &5.0    &5.0    &6.0    &26.3    &5.0   &5.0    &5.0    &6.0    &176.6 \\
			\multicolumn{11}{l}{$\rho=0.5$}\\
			DCSIS        &75.9 &257.8  &434.5  &719.3  &1167.5  &232.9 &752.8  &1230.0 &1632.5 &1936.1\\
			BCorSIS      &233.7&679.3  &1002.0 &1279.8 &1825.5  &1072.9&1405.5 &1712.0 &1888.5 &1970.1\\
			PS           &5.0  &5.0    &5.0    &5.0    &5.0     &5.0   &5.0    &24.0   &426.0  &1707.4\\
            RCC$_{sp}$   &5.0  &5.0    &5.0    &5.0    &8.1     &5.0   &5.0    &8.0    &15.0   &78.0 \\
            RCC$_{kd}$   &5.0  &6.0    &14.0   &33.3   &134.9   &5.0   &9.0    &17.5   &46.5   &204.3\\
			ELSIS$_{avg}$&59.0 &204.0  &355.0  &542.8  &906.1   &44.0  &185.8  &380.5  &654.0  &1087.8\\
			ELSIS$_{max}$&12.0 &62.3   &192.5  &451.5  &926.2   &17.9  &96.0   &321.0  &639.3  &1221.3\\
			MELSIS       &5.0  &5.0    &5.0    &5.0    &8.0     &5.0   &5.0    &5.0    &6.0    &176.6\\\bottomrule
		\end{tabular}
		}
	\end{table}

\textbf{Example 4.3.}
This experiment is used to check the effectiveness of CMELSIS when there is hidden active variables in the model. Consider the following model:
\begin{equation*}
	\left\{\begin{split}
	&Y_1 = X_1+2X_2+3X_3-3X_4+\varepsilon_1\\
	&Y_2 = 2X_1-2X_2+2X_3-3X_4+\varepsilon_2\\
	&Y_3 = X_1+2X_2+X_3-3X_4+X_5+\varepsilon_3
	\end{split}\right.
\end{equation*}
where $X_j\sim N(0,1)$ for $j=1,\cdots, p$ with equi-correlation 0.5 among predictors except that we set $X_5$ being independent of the others. By this design, $X_5$ can be seen as a hidden important variable because its marginal utility by MELSIS is almost zero. Actually, it is easily seen from the simulation results that $X_1$ is also important but is easily missed by unconditional screening methods. To check the robustness of CMELSIS to the choice of conditioning set, we consider the following different choices for the conditional set:
(1) $\mathcal{C}_1=\{2,3,4\}$; (2) $\mathcal{C}_2=\{1,2,3\}$; (3) $\mathcal{C}_3=\{1,2,10\}$;  (4) $\mathcal{C}_4=\{1, 9, 10\}$, where (1) and (2) are two ideal situations where all memberships in $\mathcal{C}$ are active, (3) contains one inactive variable and (4) only has one active variable.
The random errors follow similar settings to Example 1 that  $\varepsilon_i=\sigma_i({\bf X})\epsilon_i$ with (a) $\sigma_i(\mathbf{X})=1\mathrm{~for~} i=1,2,3$, and (b) $\sigma_1(\mathbf{X})=X_1, \sigma_2(\mathbf{X})=X_3,\sigma_3(\mathbf{X})=X_5$, respectively, and $\epsilon_i\sim N(0,1)$.
Simulation results with $(n,p)=(100,1000)$ are shown in Table \ref{ms_ex3}. From these tables, the following conclusions can be summarized:
\begin{enumerate}
	\item [(1)] CMELSIS has an excellent performance compared with its competitors CELSIS$_{avg}$ and CELSIS$_{max}$. The unconditional screening method MELSIS is only able to select $X_2, X_3$ and $X_4$ into model but misses the other two active predictors.
	\item [(2)] When the conditional set is chosen as $\mathcal{C}_1$ in which all predictors are active, all the conditional methods can successfully select the remaining active variables $X_1$ and $X_5$ into model, but our method provides a better result with smaller MMS and larger $P_a$.
	\item [(3)] When more inactive variables are added into the conditional set, our method still works well but the other two methods collapse rapidly.
	\item [(4)] Compared with the hard-threshold rule, the soft-thresholding rule sometimes results in a better performance in terms of $P_a$, but the determined model size is larger than $n/\log n$ with a large variance.
\end{enumerate}

\textbf{Example 4.4.} This experiment aims to check the effectiveness of the proposed two-step screening method when the conditional set is unavailable. We still use the model in Example 4.3 and keep all the model settings unchanged. We only present the simulation results corresponding to the case of homogeneous variance in random errors. For the heterodastic situation, the corresponding result is of course a little bit worse.  Table \ref{two-stage} reports the proportion of all the active predictors being selected under different model size $d$. Here, we set $d_n$ equal to $[n/\log n]=21$, $[1.5n/\log n]=32$ and $2[n/\log n]=42$ to represent a small, moderate and large model size, respectively.  From this table,  it is easily seen that the proposed two stage method performs very well even when the conditional set only contains three variables but the other two methods behave badly even when the conditional set contains nine variables. 

\begin{table}[htbp]
\centering
\caption{The proportion of active predictors being selected for MELSIS in Example 4.1 and CMELSIS in Example 4.3, under soft thresholding rule. The median of MMS (MMMS) associated with the interquartile range in the parenthesis are also reported.}\label{proportion_ex13_soft}
\scalebox{0.93}{
\begin{tabular}{lcccccccccc}
\toprule
&&&\multicolumn{4}{c}{$\sigma_i(\mathbf{X})$:~case (a)}&\multicolumn{4}{c}{$\sigma_i(\mathbf{X})$:~case (b)}\\
\cmidrule(lr){4-7}\cmidrule(lr){8-11}
& & &\multicolumn{2}{c}{$\tau=0.99$}&\multicolumn{2}{c}{$\tau=0.98$}&\multicolumn{2}{c}{$\tau=0.99$}&\multicolumn{2}{c}{$\tau=0.98$} \\ \cmidrule{4-11}
&$\rho$&&$P_a$& MMMS&$P_a$& MMMS&$P_a$& MMMS&$P_a$& MMMS \\ \midrule
MELSIS&0.0&&0.79&7.0(3.0)&0.92&16.0(5.5)&0.71&25.0(6.0)&0.82&43.5(12.0)\\
	  &0.5&&0.94&5.0(2.0)&0.99&15.0(5.3)&0.77&26.0(8.3)&0.89&46.5(7.5)\\
CMELSIS& &$\mathcal{C}_1$&0.99&4.0(3.0)&0.99&7.0(8.3)&0.93&5.0(3.1)&0.95&8.0(6.0)\\
		&&$\mathcal{C}_2$&0.89&19.0(36.5)&0.92&87.0(76.1)&0.83&34.0(56.3)&0.88&66.5(69.0)\\
		&&$\mathcal{C}_3$&0.88&19.5(26.5)&0.91&52.0(54.5)&0.81&26.5(28.4)&0.83&48.5(56.5)\\
		&&$\mathcal{C}_4$&0.83&27.5(55.4)&0.86&78.0(64.3)&0.76&36.5(48.5)&0.79&54.5(67.3)\\
\bottomrule
\end{tabular}
}
\end{table}

\subsection{Real data analysis}
In this section, we apply our method to a real example of genetic regulation. This data set consists of 29 inbred rats samples with a 4 dimensional response and 770 dimensional predictors. Specifically, the 4 dimensional response are a quantitative phenotype representing the expression levels of four organs, including adrenal gland, heart, kidney and fat, respectively, and the predictors are 770 single nucleotide polymorphisms.
The dataset is from a study of \cite{matthias2020a} and is available from R package R2GUESS. Our goal is to discover the genetic causes of variation in the expression of genes, i.e., to identify the SNPs that explain the joint variability of gene expression in all organs, this is the typical analysis known as expression Quantitative Trait Loci (eQTL).

Table \ref{snps} displays the top 29 ranked SNPs selected by our method and the competitors, including ELSIS$_{avg}$, ELSIS$_{max}$ and PS. It can be seen that our method selects very different SNPs compared to ELSIS$_{avg}$ or ELSIS$_{max}$ but select vary similar results to PS does. Table \ref{snps} shows that there are 20 overlapping SNPs between our method and PS but only 2 overlapping SNPs between our method and ELSIS$_{avg}$ or ELSIS$_{max}$.
It has been identified previously by \cite{matthias2020a} that the SNP D14Mit3 is a very important SNP associated with all organs. Our method successfully ranks this SNP in the top position, which strongly demonstrates the effectiveness of our method. Unfortunately, neither ELSIS$_{avg}$ nor ELSIS$_{max}$ pick out D14Mit3 as the significant SNPs.

To further check the effectiveness of our method, we in the following propose a two-stage variable selection procedure, i.e., in the first stage, we apply the screening procedure to lower the huge dimensionality $p$ to a moderate scale $s$, then in the second stage, we employ some variable selection methods such as lasso to make a further variable selection and  parameter estimation.
For this inbred dataset, we apply the MELSIS method followed by lasso to make the variable selection. Note that because the response is multivariate, we need to apply the lasso response by response. We denote this two-stage method by MELSIS---LASSO. Similarly, we can also define ELSIS$_{avg}$---LASSO, ELSIS$_{max}$---LASSO and PS---LASSO. During the application of the two-stage procedure, we set $s=[n/2]=14$ for a small model and $n =29$ for a large model, respectively, we also use the BIC criterion to determine the model size in the second variable selection stage. Table \ref{RSS} reported the corresponding residual sum of square (RSS) associated with the model size (in the parenthesis) determined by BIC, where RSS=$1/n\sum_{i=1}^n(y_i-\hat{y}_i)^2$.

From Table \ref{RSS}, we can obtain the following observations. First, it is consistent with the above analysis that the newly proposed method has a superior performance than ELSIS$_{avg}$ or ELSIS$_{max}$, the latter two methods behave badly even we set the thresholding value equal to the sample size 29, especially for the third response, kidney, lasso selects none variables into the model, this might result from that ELSIS$_{avg}$ or ELSIS$_{max}$ have a completely wrong result. Besides, we see that MELSIS behaves better than PS fo1r the 1st and the last response but worse than PS for the remaining two responses.

By the way, we also apply the newly proposed conditional screening procedure to the inbred dataset. As the SNPs D14Mit3, D14Cebrp312s2 and D14Rat52 are three common SNPs selected by MELSIS and PS, we then set the three SNPs as the prior information when implementing CMELSIS, the corresponding numerical results are still presented in Table \ref{RSS},  from which we see that CMELSIS does improve the performance compared to the unconditional methods MELSIS except for the last response.

\begin{table}[H]
\centering
\caption{RSS of different methods and the associated model size(in the parenthesis). Resp 1, 2, 3 and 4 corresponds to the adrenal gland, heart, kidney and fat, respectively.}\label{RSS}
\begin{tabular*}{\hsize}{@{}@{\extracolsep{\fill}}lllllll@{}}
\toprule
$s$ & Method & Resp 1 & Resp 2 & Resp 3 & Resp 4 \\ \midrule
14 &MELSIS---LASSO        &0.0434(2)  &0.0577(2)  &0.2943(1) &0.0537(1) \\
   &ELSIS$_{avg}$---LASSO &0.0646(0)  &0.0610(2)  &1.5530(0) &0.0604(1) \\
   &ELSIS$_{max}$---LASSO &0.0646(0)  &0.0610(2)  &1.5530(0) &0.0604(1) \\
   &PS---LASSO            &0.0414(1)  &0.0373(2)  &0.2142(2) &0.0358(1) \\
   &RCC$_{sp}$-LASSO      &0.0414(2)  &0.0468(3)  &0.2943(1) &0.0358(2) \\
   &RCC$_{kd}$-LASSO      &0.0382(3)  &0.0506(3)  &0.2943(1) &0.0358(2) \\
   &CMELSIS---LASSO       &0.0313(2)  &0.0376(4)  &0.2877(1) &0.0508(1) \\ \midrule
29 &MELSIS---LASSO        &0.0079(13) &0.0467(4)  &0.1693(4) &0.0166(9) \\
   &ELSIS$_{avg}$---LASSO &0.0483(2)  &0.0355(6)  &1.5530(0) &0.0604(1) \\
   &ELSIS$_{max}$---LASSO &0.0483(2)  &0.0355(6)  &1.5530(0) &0.0604(1) \\
   &PS---LASSO            &0.0147(8)  &0.0112(10) &0.0890(8) &0.0187(6) \\
   &RCC$_{sp}$-LASSO      &0.0087(13) &0.0299(4)  &0.1587(3) &0.0358(2) \\
   &RCC$_{kd}$-LASSO      &0.0095(12) &0.0140(9)  &0.2142(3) &0.0095(11) \\
   &CMELSIS---LASSO       &0.0066(14) &0.0041(13) &0.1367(4) &0.0306(1) \\\bottomrule
\end{tabular*}	
\end{table}

\section*{Acknowledgement}
Jun Lu's research was partly supported by National Natural Science Foundation (NNSF) of China (No.12001486). Qinqin Hu's research was supported by NNSF of China (No. 11601283). Lu Lin's research  was supported by NNSF of China (No. 11971265).

\appendix
\section{Proofs}

\textbf{Proof of Proposition 3.1.} Let ${\bf B}_\A=(\bm\beta_j: j\in\A)$ be the coefficient matrix corresponding to the active predictors and $K=|\A|$. Without loss of generality, assume that $\bm\beta_j^\top\mbox{cov}(\X_\A,\X_\A^\top)\bm\beta_j=1$ for $j\in \A$, otherwise, we can divide $\bm\beta_j$ by a constant such that this condition is satisfied. It has that 
\begin{eqnarray*}
&&\max_{j\in\I}\|\E X_j\y\|^2 = \max_{j\in\I}\|\E{\bf B}_\A \X_\A X_j \|^2 
\leq \max_{j\in\I}\sum_{k\in\A} |\bm\beta_k^\top\E(\X_\A\X_j)|^2 \\
&\leq & \sum_{k\in\A}^K \|\bm\beta_k^\top\E(\X_\A\X_\I^\top)\|^2 
=\sum_{k\in\A}\bm\beta_k^\top\mbox{cov}(\X_\A,\X_\I^\top)\mbox{cov}(\X_\I,\X_\A^\top)\bm\beta_k\\
&\leq & \sum_{k\in\A}\frac{ \bm\beta_k^\top\mbox{cov}(\X_\A,\X_\I^\top)\mbox{cov}(\X_\I,\X_\A^\top)\bm\beta_k}{\bm\beta_k^\top\mbox{cov}(\X_\A,\X_\A^\top)\bm\beta_k}\\
&\leq & \frac{K\lambda_{\max}(\mbox{cov}(\X_\A,\X_\I^\top)\mbox{cov}(\X_\I,\X_\A^\top))}
{\lambda_{\min} (\mbox{cov}(\X_\A,\X_\A^\top))}
\end{eqnarray*}
\qed

\textbf{Proof of Proposition 3.2.} It has that 
\begin{eqnarray*}
&&\max_{j\in\I}\|\E\wXj\y\|^2 = \max_{j\in\I}\|\E{\bf B}_\A \X_\A (X_j-\E(X_j|\BC^\top\XC)) \|^2 \\
&=&\max_{j\in\I}\|\E({\bf B}_\A \X_\A X_j)-\E\left\{{\bf B}_\A \X_\A\XC^\top \BC\right\}\mathrm{cov}^{-1}(\BC^\top\XC)\E(\BC^\top\XC, X_j) \|^2 \\
&=&\max_{j\in\I}\|\E({\bf B}_\A \X_\A X_j)-\E (X_j \E({\bf B}_\A\X_\A|\BC^\top\XC) )\|^2  \\ 
&\leq & \max_{j\in\I}\sum_{k\in\A} |\bm\beta_k^\top\E(\widetilde{\X}_\A\X_j)|^2 \\
&\leq & \frac{K\lambda_{\max}(\mbox{cov}(\widetilde{\X}_\A,\X_\I^\top)\mbox{cov}(\X_\I,\widetilde{\X}_\A^\top))}
{\lambda_{\min} (\mbox{cov}(\widetilde{\X}_\A,\widetilde{\X}_\A^\top))},
\end{eqnarray*}
where the first and second equalities hold because of the linear condition. 
\qed

\textbf{Proof of Lemma~4.1}
Define $U_{i,jk}=X_{ij}Y_{ik}$ and $\mu_{jk}=\E(U_{i,jk})$. By Cauchy–Schwarz inequality, it has that $|\mu_{jk}|\leq (\E(X_{ij}^2))^{1/2}(\E(Y_{ik}^2))^{1/2}$, then $|\mu_{jk}|$ can be bounded by a uniform constant.
 Without loss of generality, we assume that $\mu_{jk}>0$. If $\mu_{jk}<0$, we can let $\widetilde {U}_{i,jk}=-U_{i,jk}$. Note that
\begin{align*}
EL_{j}(\bm{0})&=\sup \left\{\prod_{i=1}^{n} w_{i} : w_{i} \geq 0, \sum_{i=1}^{n} w_{i}=1, \sum_{i=1}^{n} w_{i} \bm {u}_{i j}=0\right\}\\
&=\sup \left\{\prod_{i=1}^{n} w_{i} : w_{i} \geq 0, \sum_{i=1}^{n} w_{i}=1, \sum_{i=1}^{n} w_{i} \bm {\tilde u}_{i j}=0\right\},
\end{align*}
where $\bm {u}_{i j}=(U_{i,j1},\ldots,U_{i,jq})^\top$ and $\tilde{\bm u}_{i j}=(\widetilde U_{i,j1},\ldots,\widetilde U_{i,jq})^\top$. Hence $$l_{j}(\bm{0})=-2 \log \left\{\mathrm{EL}_{j}(\bm{0})\right\}-2 n \log n$$ does not depend on the sign of $\mu_{jk}$.

For given $j\in \A$, according to \cite{owen2001empirical}, we have that $$l_j(\bm 0)=2 \max _{\bm \alpha \in \Lambda_{n,j}} \sum_{i=1}^{n} \log \left(1+\bm \alpha^\top \bm { u}_{i j}\right),$$ where
$\Lambda_{n, j}=\left\{\bm \alpha : \bm \alpha^\top \bm { u}_{i j} \geq n^{-1}\text{ for all\,} i=1, \ldots, n\right\}$.

Set $\bm\alpha=\bm a=\left(n^{\epsilon} \max _{i,k}\left|U_{i,jk}\right|\right)^{-1}(1,\ldots,1)^\top$ for some $\epsilon>0$, then $\bm a\in \Lambda_{n, j}$ for sufficiently large $n$. Hence
$$\mathbb{P}\{l_j(\bm 0)<2 t\} \leq \mathbb{P}\left\{\sum_{i=1}^{n} \log \left[1+\sum_{k=1}^q\frac{U_{i,jk}}{n^{\epsilon} \max _{i,k}\left|U_{i,jk}\right|}\right]<t\right\}.$$
By Taylor expansion,
$$
\log\left[1+\sum_{k=1}^q\frac{U_{i,jk}}{n^{\epsilon} \max _{i,k}\left|U_{i,jk}\right|}\right]=\frac{\sum_{k=1}^q U_{i, jk}}{n^{\epsilon} \max _{i,k}\left|U_{i,jk}\right|}-\frac{1}{2\left(1+c_{i}\right)^{2}} \frac{(\sum_{k=1}^q U_{i, jk})^{2}}{n^{2\epsilon} \max _{i,k}\left|U_{i,jk}\right|^2},
$$
where $\left|c_{i}\right| \leq n^{-\epsilon}$, then
$$
\sum_{i=1}^{n} \log\left[1+\sum_{k=1}^q\frac{U_{i,jk}}{n^{\epsilon} \max _{i,k}\left|U_{i,jk}\right|}\right]=\sum_{i=1}^{n} \frac{ \sum_{k=1}^qU_{i, jk}}{n^{\epsilon} \max _{i,k}\left|U_{i,jk}\right|}+R_{n}
$$
with $\left|R_{n}\right| \leq n^{1-2 \epsilon}$. Therefore,
$$
\mathbb{P}\{l_j(\bm 0)<2 t\} \leq \mathbb{P}\left\{\sum_{i=1}^{n} \frac{\sum_{k=1}^q U_{i, jk}}{n^{\epsilon} \max _{i,k}\left|U_{i,jk}\right|}<t+n^{1-2 \epsilon}\right\},
$$
which means that
\begin{align*}
&\quad~ \mathbb{P}\{l_j(\bm 0)<2 t\} \\
&\leq \mathbb{P}\left\{\sum_{i=1}^{n}\sum_{k=1}^q U_{i, jk}<(t n^{\epsilon}+n^{1-\epsilon}) \max _{i,k}\left|U_{i,jk}\right|\right\}\\
&\leq \mathbb{P}\left\{\sum_{i=1}^{n}\sum_{k=1}^q U_{i, jk}-n\sum_{k=1}^q \mu_{jk}<(t n^{\epsilon}+n^{1-\epsilon})M -n\sum_{k=1}^q \mu_{jk}\right\}+\mathbb{P}\left\{ \max _{i,k}\left|U_{i,jk}\right|>M\right\}\\
&\leq \sum_{k=1}^q \mathbb{P}\left\{\sum_{i=1}^{n} (U_{i, jk}-\mu_{jk})<\frac{(t n^{\epsilon}+n^{1-\epsilon})M-n\sum_{k=1}^q \mu_{jk}}{q} \right\}+\mathbb{P}\left\{ \max _{i,k}\left|U_{i,jk}\right|>M\right\}\\
&=\sum_{k=1}^q \mathbb{P}\left\{\frac{1}{\sqrt{n}\sigma_{jk}}\sum_{i=1}^{n} (U_{i, jk}-\mu_{jk})<\frac{(t n^{\epsilon-\frac{1}{2}}+n^{\frac{1}{2}-\epsilon})M-n^\frac{1}{2}\sum_{k=1}^q \mu_{jk}}{q\sigma_{jk}} \right\}\\
&\quad +\mathbb{P}\left\{ \max _{i,k}\left|U_{i,jk}\right|>M\right\},
\end{align*}
where $\sigma_{jk}^2=\E(U_{i,jk}-\mu_{jk})^2$. For $L \rightarrow \infty$, pick $\epsilon$ satisfies $n^{\epsilon}=L/({\sum_{k=1}^q \mu_{jk}})$. Choose $\eta \in\left(0, \frac{1}{2}\right)$, let $M=\eta L$ and $2 t={n (\sum_{k=1}^q \mu_{jk})^{2}}/L^{2},$ then ${t n^{\epsilon} M}/({n \sum_{k=1}^q \mu_{jk}})={\eta}/{2}$ and ${n^{1-\epsilon} M}/({n \sum_{k=1}^q \mu_{jk}})=\eta$. Hence, for sufficient large $n$, condition (C2) together with Lemma 1 of \cite{chang2013marginal} lead to
{
\begin{align*}
&\quad~\mathbb{P}\left\{l_j(\bm 0)<\frac{c_1^2n^{1-2\kappa}}{L^2}\right\}\leq \mathbb{P}\left\{l_j(\bm 0)<\frac{n\|\mathbb{E}\{X_j{\bf y}\}\|_2^2}{L^2}\right\}\leq\mathbb{P}\left\{l_j(\bm 0)<\frac{n(\sum_{k=1}^q\mu_{jk})^2}{L^2}\right\}\\
&\leq\sum_{k=1}^q \mathbb{P}\left\{\frac{1}{n^\frac{1}{2}\sigma_{jk}}\sum_{i=1}^{n} (U_{i, jk}-\mu_{jk})<\frac{(\frac{3}{2}\eta-1)n^\frac{1}{2}\sum_{k=1}^q \mu_{jk}}{q\sigma_{jk}} \right\}+K_{1} \exp \left\{-K_{2} M^{\gamma}+\log n+\log q\right\}\\
&\leq \left\{\begin{array}{c}{\exp \left(-C_1 n^{1-2 \kappa-2r}\right)+\exp \left(-C_1 L^{\gamma}\right), \text { if }(1-2 \kappa-2r)(1+2\delta)<1} \\ {\exp \left(-C_1 n^{\frac{1-\kappa-r}{1+\delta}}\right)+\exp \left(-C_1 L^{\gamma}\right), \text { if }(1-2 \kappa-2r)(1+2\delta)\geq 1}\end{array}\right.
\end{align*}
}

\qed

\textbf{Proof of Theorem~4.1}.
Note that
\begin{align*}
\mathbb{P}\left\{\A\varsubsetneq \widehat{\A}_{\gamma_{n}}\right\}&=\mathbb{P}\left\{ \,\text{There exists} \,j \in \A \,\text{such that} \,\ell_j(\bm 0)<c_{1}^{2} n^{2 \tau}\right\}\\
&\leq s \max _{j \in \A} \mathbb{P}\left\{\ell_j(\bm 0)<c_{1}^{2} n^{2 \tau}\right\},
\end{align*}
then we can get our result directly by Lemma~2.1.
\qed

\textbf{Proof of Theorem~4.2.} Before proving Theorem~4.2. We first prove Lemma A1 below. 

{\sc Lemma A1.} 
Let $\bm\theta=\bm\alpha/\|\bm\alpha\|_2$, if $\max_{j,k}|\mu_{jk}|= \max_{j,k}|\E(X_{j}Y_{k})|=O_p(n^{-\eta})$ for some $\eta>0$, then

{
\begin{align*}
&\mathbb{P}\left\{\|\bm\alpha\|_2>\frac{4\bm\theta^\top \bar{\bm u}_{j}}{3\bm\theta^\top\bm S_j\bm\theta}\right\} \leq \\
&\qquad \left\{\begin{array}{l}{\exp \left(-C n^{\gamma (\eta-3r/2)\wedge\gamma(1-\eta)/2\wedge(1-4r)}\right), \text { if } \gamma < 2 \text { and } r \geq \delta_2 \text{and }  \eta<\delta_1} 
\\ {\exp \left(-C n^{\gamma (\eta-3r/2)\wedge(1-2\eta)\wedge(1-r+\eta)\gamma/2\wedge(1-4r)}  \right), } {\text { if } \gamma < 2 \text { and } r \geq \delta_2\text{ and } \eta\geq \delta_1}
\\ {\exp \left(-C n^{\gamma (\eta-3r/2)\wedge(1-2\eta)\wedge(1-2r)\gamma/4}\right),}  {\text { if } \gamma < 2 \text { and } r< \delta_2  \text{ and }  \eta\geq \delta_1}\\ 
{\exp \left(-C n^{\gamma (\eta-3r/2)\wedge(1-\eta)\gamma/2\wedge(1-2r)\gamma/4}\right), \text { if } \gamma < 2 \text { and } r< \delta_2 \text{ and } \eta<\delta_1}\\ 
{\exp \left(-C n^{\gamma (\eta-3r/2)\wedge(1-2\eta)\wedge(1-4r)}\right), \text { if } 2\leq\gamma <4 \text { and } r\geq \delta_2}\\ 
{\exp \left(-C n^{\gamma (\eta-3r/2)\wedge(1-2\eta)\wedge(1-2r)\gamma/4}\right), \text { if } 2\leq\gamma <4 \text { and } r<\delta_2}\\ {\exp \left(-C n^{\gamma (\eta-3r/2)\wedge(1-2\eta)\wedge(1-4r)}\right), \text { if } \gamma\geq 4}\end{array}\right.
\end{align*}}
where $\bar{\bm u}_{j}=\frac{1}{n}\sum_{i=1}^n\bm u_{i j}$, $\bm S_j=\frac{1}{n}\sum_{i=1}^n\bm u_{ij}\bm u_{ij}^\top $, $\bm \alpha$ satisfies the equation $\bm 0=\sum_{i=1}^{n} \frac{\bm u_{i j}}{1+\bm\alpha^\top  \bm u_{i j}}$, $\delta_1=\frac{1}{2}-\frac{\gamma}{8-2\gamma}$ and $\delta_2=\frac{1}{4}-\frac{\gamma}{32-4\gamma}$, $C$ is some positive constant depending on $K_1, K_2,\gamma_1$ and $\gamma_2$ given in Condition (C2).

\proof
{
Since $\bm\alpha$ satisfies the equation $\sum_{i=1}^n\frac{\bm u_{ij}}{1+\bm\alpha^\top\bm u_{ij}}=\bm0$, thus it can be deduced that 
$$\bm0=\sum_{i=1}^n\frac{\bm u_{ij}}{1+\bm\alpha^\top\bm u_{ij}}=\sum_{i=1}^n\bm u_{ij}-\sum_{i=1}^n\frac{\bm u_{ij}\bm u_{ij}^\top \bm\alpha}{1+\bm\alpha^\top\bm u_{ij}}.$$
Define $\bm\theta=\bm\alpha/\|\bm\alpha\|_2$, it is easily proved that 
$$\bm\theta^\top\frac1n\sum_{i=1}^n\bm u_{ij}=\bm\theta^\top\frac1n\sum_{i=1}^n\frac{\bm u_{ij}\bm u_{ij}^\top}{1+\bm\alpha^\top\bm u_{ij}} \bm\alpha=\bm\theta^\top\frac1n\sum_{i=1}^n\frac{\bm u_{ij}\bm u_{ij}^\top}{1+\bm\alpha^\top\bm u_{ij}} \bm\theta\|\bm\alpha\|_2.$$
Let $\bar{\bm u}_{j}\frac1n\sum_{i=1}^n\bm u_{ij}$, $\widetilde{\bm S}_j=\frac{1}{n}\sum_{i=1}^n\frac{\bm u_{ij}\bm u_{ij}^\top }{1+\bm \alpha^\top  \bm u_{i j}}$, we then have  
$$\|\bm \alpha\|_2=\frac{\bm\theta^\top \bar{\bm u}_{j}}{\bm\theta^\top\widetilde{\bm S}_j\bm\theta}.$$
}
Because $\max_l\bm \alpha^\top  \bm u_{l j}\leq \|\bm \alpha\|_2\max_{l}\|\bm u_{l j}\|_2$, according the similar argument in Lemma 4 in \cite{chang2013bsupplement}, we have
$$\mathbb{P}\left\{\|\bm \alpha\|_2<\frac{4\bm\theta^\top \bar{\bm u}_{j}}{3\bm\theta^\top \bm S_j\bm\theta}\right\}\geq \mathbb{P}\left\{\frac{\bm\theta^\top \bar{\bm u}_{j}}{\bm\theta^\top \bm S_j\bm\theta}\max_{l}\|\bm u_{l j}\|_2<\frac{1}{4}\right\}. $$

Pick 
{ $e\in(0,\eta-r)$}, then
\begin{align*}
 & \mathbb{P}\left\{\frac{\bm\theta^\top \bar{\bm u}_{j}}{\bm\theta^\top\bm S_j\bm\theta}\max_{l}\|\bm u_{l j}\|_2\geq\frac{1}{4}\right\} \\
\leq  ~&\mathbb{P}\left\{\max_{l}\|\bm u_{l j}\|_2\geq\frac{n^e}{4}\right\}+\mathbb{P}\left\{\bm\theta^\top \bar{\bm u}_{j}\geq n^{-e}\bm\theta^\top\bm S_j\bm\theta\right\}\\
\leq  ~&\mathbb{P}\left\{\max_{l}\|\bm u_{l j}\|_2\geq\frac{n^e}{4}\right\}+\mathbb{P}\left\{\theta^\top \bar{\bm u}_{j}\geq n^{-e}\lambda_{min}\right\}+\mathbb{P}\left\{\bm\theta^\top \bm S_j\bm\theta<\lambda_{min}\right\}\\
=:~&I_1+I_2+I_3,
\end{align*}
where $\lambda_{min}$ is the smallest eigenvalue of $Var(\bm u_{ij})$. 

We next bound the three items one by one. The first item $I_1$ can be easily bounded as $I_1\leq \exp\{-Cn^{(e-r/2)\gamma}\}$. For the second iterm $I_2$,  according to Lemma 1 in \cite{chang2013bsupplement},  we can get
\begin{align*}
 I_2&=\mathbb{P}\left\{\bm\theta^\top \bar{\bm u}_{j}\geq n^{-e}\lambda_{min}\right\}\leq q\mathbb{P}\left\{\sum_{i=1}^n\frac{1}{n}U_{i,jk}\geq \frac{ n^{-e}\lambda_{min}}{q\theta_k}\right\}\\
 &\leq  \left\{\begin{array}{c}{n^{r}\exp \left\{-Cn^{1-2r-2e}\right\}, \text { if } (1-2r-2e)(1+2\delta)<1 }\\ {n^{r}\exp \left\{-Cn^{\frac{1-r-e}{1+\delta}}\right\}, \text { if } (1-2r-2e)(1+2\delta)>1}\end{array}\right. .
\end{align*}
For the third item, let $\theta_{kt}=\theta_k\theta_t$, then
\begin{align*}
 I_3&=\mathbb{P}\left\{\bm\theta^\top\bm S_j\bm\theta<\lambda_{min}\right\}\\
 &=\mathbb{P}\left\{\sum_{k,t=1}^q\sum_{i=1}^n\frac{1}{n}U_{i,jk}U_{i,jt}\theta_{kt}<\lambda_{min}\right\}\\
 &=\mathbb{P}\left\{\sum_{k,t=1}^q\sum_{i=1}^n\frac{1}{n}[(U_{i,jk}-\mu_{jk})(U_{i,jt}-\mu_{jt})-\sigma^2_{i,jkt}]\theta_{kt}+\Upsilon<\lambda_{min}-\bm\theta^\top Var(\bm u_{ij})\bm\theta\right\}\\
 &\leq\mathbb{P}\left\{\sum_{k,t=1}^q\sum_{i=1}^n\frac{1}{n}[(U_{i,jk}-\mu_{jk})(U_{i,jt}-\mu_{jt})-\sigma^2_{i,jkt}]\theta_{kt}<\frac{\lambda_{min}-\bm\theta^\top Var(\bm u_{ij})\bm\theta}{2}\right\}\\
 &+\mathbb{P}\left\{\Upsilon<\frac{\lambda_{min}-\bm\theta^\top Var(\bm u_{ij})\bm\theta}{2}\right\}\\
 &:=I_{31}+I_{32},
\end{align*}
where $\Upsilon=\sum_{k,t=1}^q\sum_{i=1}^n\frac{1}{n}(\mu_{jk}U_{i,jt}\theta_{kt}+\mu_{jt}U_{i,jk}\theta_{tk}-\theta_{kt}\mu_k\mu_t)$ and $Var(\bm u_{ij})=(\sigma^2_{i,jkt})_{q\times q}$.

By Lemma 1 in \cite{chang2013bsupplement}, $I_{31}$ can be bounded as 
\begin{align*}
 I_{31}&\leq\sum_{k,t=1}^q\mathbb{P}\left\{\sum_{i=1}^n\frac{1}{n}[(U_{i,jk}-\mu_{jk})(U_{i,jt}-\mu_{jt})-\sigma^2_{i,jkt}]\theta_{kt}<\frac{\lambda_{min}-\bm\theta^\top Var(\bm u_{ij})\bm\theta}{2q^2}\right\}\\
 &\leq \left\{\begin{array}{c}{n^{2r}\exp \left\{-Cn^{1-4r}\right\}, \text { if } (1-4r)(1+2\tilde\delta)<1 }\\ {n^{2r}\exp \left\{-Cn^{\frac{1-2r}{1+\tilde\delta}}\right\}, \text { if } (1-4r)(1+2\tilde\delta)>1}\end{array}\right.
\end{align*}
where $q=O(n^r)$ and $\tilde\delta=\max\{\frac{4}{\gamma}-1,0\}$. Similarly, $I_{32}$ is bounded as
\begin{align*}
 I_{32}\leq \left\{\begin{array}{c}{n^{r}\exp \left\{-Cn^{1-2r+2\eta}\right\}, \text { if } (1-2r+2\eta)(1+2\delta)<1 }\\ {n^{r}\exp \left\{-Cn^{\frac{1-r+\eta}{1+\delta}}\right\}, \text { if } (1-2r+2\eta)(1+2\delta)>1}\end{array}\right.
\end{align*}

Finally, let  $\delta_1=\frac{1}{2}-\frac{\gamma}{8-2\gamma}$ and $\delta_2=\frac{1}{4}-\frac{\gamma}{32-4\gamma}$, combing the above several results, we can obtain that
{
\begin{align*}
&\mathbb{P}\left\{\frac{\bm\theta^\top \bar{\bm u}_{j}}{\bm\theta^\top\bm S_j\bm\theta}\max_{l}\|\bm u_{l j}\|_2\geq\frac{1}{4}\right\} \leq \\
&\qquad \left\{\begin{array}{l}{\exp \left(-C n^{\gamma (\eta-3r/2)\wedge\gamma(1-\eta)/2\wedge(1-4r)}\right), \text { if } \gamma < 2 \text { and } r \geq \delta_2 \text{ and }  \eta<\delta_1} 
\\ {\exp \left(-C n^{\gamma (\eta-3r/2)\wedge(1-2\eta)\wedge(1-r+\eta)\gamma/2\wedge(1-4r)}  \right), } {\text { if } \gamma < 2 \text { and } r\geq \delta_2 \text{ and } \eta\geq \delta_1}
\\ {\exp \left(-C n^{\gamma (\eta-3r/2)\wedge(1-2\eta)\wedge(1-2r)\gamma/4}\right),}  {\text { if } \gamma < 2 \text { and } r< \delta_2  \text{ and }  \eta\geq \delta_1}\\ 
{\exp \left(-C n^{\gamma (\eta-3r/2)\wedge(1-\eta)\gamma/2\wedge(1-2r)\gamma/4}\right), \text { if } \gamma < 2 \text { and } r< \delta_2 \text{ and } \eta<\delta_1}\\ 
{\exp \left(-C n^{\gamma (\eta-3r/2)\wedge(1-2\eta)\wedge(1-4r)}\right), \text { if } 2\leq\gamma <4 \text { and } r\geq \delta_2}\\ 
{\exp \left(-C n^{\gamma (\eta-3r/2)\wedge(1-2\eta)\wedge(1-2r)\gamma/4}\right), \text { if } 2\leq\gamma <4 \text { and } r<\delta_2}\\ {\exp \left(-C n^{\gamma (\eta-3r/2)\wedge(1-2\eta)\wedge(1-4r)}\right), \text { if } \gamma\geq 4}\end{array}\right.
\end{align*}
}
The lemma is proved. 
\qed

Now we turn to the proof of Theorem 4.2. It has that
\begin{align*}
|\widehat{\A}_{\gamma_{n}}|&=\sum_{j \in \A} I\left\{l_j(\bm 0)\geq c_{1}^{2} n^{2 \tau}\right\}+\sum_{j \notin \A } I\left\{l_j(\bm 0) \geq c_{1}^{2} n^{2 \tau}\right\}\\
&\leq s+\sum_{j \notin\A} I\left\{l_j(\bm 0) \geq c_{1}^{2} n^{2 \tau}\right\},
\end{align*}
then
$$
\mathbb{P}\left\{|\widehat{\A}_{\gamma_{n}}|>s\right\} \leq \sum_{j \notin \A} \mathbb{P}\left\{l_j(\bm 0) \geq c_{1}^{2} n^{2 \tau}\right\}.
$$
Hence, it is sufficient to figure out the behavior of $\mathbb{P}\left\{l_j(\bm 0) \geq c_{1}^{2} n^{2 \tau}\right\}$ for each $j\notin \A$. 
As $\bm\theta=\bm \alpha/\|\bm \alpha\|_2$ is a unit vector, based on the vector empirical likelihood theorem, namely, Theorem 3.2 in \cite{owen2001empirical}, we can prove that
\begin{align*}
l_j(\bm 0) &=n \bar{\bm u}_{j}^\top \bm S_j^{-1}\bar{\bm u}_{j}-n\bm b_j^\top\bm S_j^{-1}\bm b_j+2\sum_{i=1}^n\eta_{ij}\\
&=:I_1+I_2+I_3,
\end{align*}
where  $\bar{\bm u}_{j}=\frac{1}{n}\sum_{i=1}^n\bm u_{i j}$, $\bm b_j=\bm S_j^{-1}\frac{1}{n}\sum_{i=1}^n\frac{\bm u_{ij}(\bm \alpha^\top  \bm u_{i j})^2}{1-\bm \alpha^\top  \bm u_{i j}}$ and $\eta_{ij}=\frac{(\bm \alpha^\top  \bm u_{i j})^3}{3(1+c_i\bm \alpha^\top  \bm u_{i j})^3}$.

If $\|\bm \alpha\|_2<\frac{4\bm\theta^\top \bar{\bm u}_{j}}{3\bm\theta^\top\bm S_j\bm\theta}$, then $\max_{l}|\bm \alpha^\top \bm u_{l j}|<\frac{4\bm\theta^\top \bar{\bm u}_{j}}{3\bm\theta^\top\bm S_j\bm\theta}\max_{l}\|\bm u_{l j}\|_2.$ Further, if $\bm\theta^\top \bar{\bm u}_{j}\max_{l}\|\bm u_{l j}\|_2<\frac{1}{4}\bm\theta^\top\bm S_j\bm\theta$, then $\max_{l}|\bm \alpha^\top \bm u_{l j}|<\frac{1}{3}$.
Define an event $$\mathcal{M}=\left\{\|\bm \alpha\|_2<\frac{4\bm\theta^\top \bar{\bm u}_{j}}{3\bm\theta^\top\bm S_j\bm\theta} \text{~and~} \bm\theta^\top \bar{\bm u}_{j}\max_{l}\|\bm u_{l j}\|_2<\frac{1}{4}\bm\theta^\top\bm S_j\bm\theta\right\},$$
then by Lemma~A1, we have
{
$$\mathbb{P}\left(\mathcal{M}^{c}\right)\leq  \left\{\begin{array}{l}{\exp \left(-C n^{\gamma (\eta-3r/2)\wedge\gamma(1-\eta)/2\wedge(1-4r)}\right), \text { if } \gamma < 2 \text { and } r \geq \delta_2 \text{ and }  \eta<\delta_1} 
\\ {\exp \left(-C n^{\gamma (\eta-3r/2)\wedge(1-2\eta)\wedge(1-r+\eta)\gamma/2\wedge(1-4r)}  \right), } {\text { if } \gamma < 2 \text { and } r\geq \delta_2 \text{ and } \eta\geq \delta_1}
\\ {\exp \left(-C n^{\gamma (\eta-3r/2)\wedge(1-2\eta)\wedge(1-2r)\gamma/4}\right),}  {\text { if } \gamma < 2 \text { and } r< \delta_2  \text{ and }  \eta\geq \delta_1}\\ 
{\exp \left(-C n^{\gamma (\eta-3r/2)\wedge(1-\eta)\gamma/2\wedge(1-2r)\gamma/4}\right), \text { if } \gamma < 2 \text { and } r< \delta_2 \text{ and } \eta<\delta_1}\\ 
{\exp \left(-C n^{\gamma (\eta-3r/2)\wedge(1-2\eta)\wedge(1-4r)}\right), \text { if } 2\leq\gamma <4 \text { and } r\geq \delta_2}\\ 
{\exp \left(-C n^{\gamma (\eta-3r/2)\wedge(1-2\eta)\wedge(1-2r)\gamma/4}\right), \text { if } 2\leq\gamma <4 \text { and } r<\delta_2}\\ {\exp \left(-C n^{\gamma (\eta-3r/2)\wedge(1-2\eta)\wedge(1-4r)}\right), \text { if } \gamma\geq 4}\end{array}\right.$$
}
If $\mathcal{M}$ holds, $$
\left|I_{3}\right| \leq C\left(n \max_\ell\left\|\bm u_{\ell j}\right\|_2^{3}\right)\left|\bm\theta^\top \bar{\bm u}_{j}\right|^{3}\left(\bm\theta^\top\bm S_j\bm\theta\right)^{-3}.
$$

Consequently, when setting{ $\breve{e}\in(0,\eta+2 \tau/3-1/3-r)$}, it has that 
{
\begin{align*}
 &\mathbb{P}\left\{I_{3} \geq \frac{c_{1}^{2} n^{2 \tau}}{2},\mathcal{M} \text { holds }\right\}\\
\leq~ &\mathbb{P}\left\{\bm\theta^\top\bm S_j\bm\theta<\lambda_{\min}\right\}+\mathbb{P}\left\{\bm\theta^\top \bar{\bm u}_{j} \geq \tilde c_{1}n^{2 \tau/3-1/3-\breve{e}}\right\}+\mathbb{P}\left\{\max_{l}\|\bm u_{lj}\|_2\geq n^{\breve{e}}\right\}.
\end{align*}
}
{
Besides, it can be proved that 
\begin{align*}
 \mathbb{P}\left\{I_{1} \geq \frac{c_{1}^{2} n^{2 \tau}}{2}\right\}&=\mathbb{P}\left\{n \bar{\bm u}_{j}^\top S_j^{-1}\bar{\bm u}_{j}\geq \frac{c_{1}^{2} n^{2 \tau}}{2}\right\}\\
 &\leq  \mathbb{P}\left\{\bar{\bm u}_{j}^\top \bar{\bm u}_{j} \geq \frac{\lambda_{\min}(\bm S_j)c_{1}^{2} n^{2 \tau-1}}{2}\right\}\\
 &= \mathbb{P}\left\{\bar{\bm u}_{j}^\top \bar{\bm u}_{j} \geq \frac{\lambda_{\min}(\bm S_j)c_{1}^{2} n^{2 \tau-1}}{2}, \lambda_{\min}(\bm S_j)\leq\lambda_{\min}\right\}\\
 &\quad + \mathbb{P}\left\{\bar{\bm u}_{j}^\top \bar{\bm u}_{j} \geq \frac{\lambda_{\min}(\bm S_j)c_{1}^{2} n^{2 \tau-1}}{2}, \lambda_{\min}(\bm S_j)>\lambda_{\min}\right\}\\
 &\leq \mathbb{P}\left\{\lambda_{\min}(\bm S_j)\leq \lambda_{\min}\right\}+\mathbb{P}\left\{\bar{\bm u}_{j}^\top \bar{\bm u}_{j} \geq \frac{\lambda_{\min}c_{1}^{2} n^{2 \tau-1}}{2}\right\} \\
&\leq \mathbb{P}\left\{\lambda_{\min}(\bm S_j)\leq \lambda_{\min}\right\}+q\mathbb{P}\left\{\left(\frac{1}{n}\sum_{i=1}^n U_{i, jk}\right)^2\geq \frac{\lambda_{\min}c_{1}^{2} n^{2\tau-1}}{2q}\right\}\\
&\leq\mathbb{P}\left\{\lambda_{\min}(\bm S_j)\leq \lambda_{\min}\right\}+ \left\{\begin{array}{l}{\exp \left(-C n^{2\tau-r}\right), \text { if } (2\tau-r)(1+2\delta)\leq 1}\\ {\exp \left(-C n^{\frac{2\tau-r+1}{2(1+\delta)}}\right), \text { if } (2\tau-r)(1+2\delta)> 1}\end{array}\right..
\end{align*}
In addtion,
\begin{align*}
	\mathbb{P}\left\{\lambda_{\min}(\bm S_j)\leq \lambda_{\min}\right\}
	=\mathbb{P}\left(\tilde{\bm\theta}\bm S_j\tilde{\bm\theta}<\lambda_{\min}\tilde{\bm\theta}^\top\tilde{\bm\theta}\right),
\end{align*}
where $\tilde{\bm\theta}$ is the corresponding eigenvector of $\lambda_{\min}(\bm S_j)$.  According to the similar analytics in the proof of Lemma A.1, we can get $\mathbb{P}\left(\tilde{\bm\theta}\bm S_j\tilde{\bm\theta}<\lambda_{\min}\tilde{\bm\theta}^\top\tilde{\bm\theta}\right)$ has the same upper bound as $\mathbb{P}\left\{\bm\theta^\top\bm S_j\bm\theta<\lambda_{\min}\right\}$.
}

Combining the results for $\mathbb{P}\left\{I_{1} \geq \frac{c_{1}^{2} n^{2 \tau}}{2}\right\}$, $\mathbb{P}\left\{I_{3} \geq \frac{c_{1}^{2} n^{2 \tau}}{2},\mathcal{M} \text { holds }\right\}$ and $\mathbb{P}\left(\mathcal{M}^{c}\right)$, together with Lemma A1, it can be easily proved that 
\begin{align*}
\mathbb{P}\left\{l_{j}(\bm 0) \geq c_{1}^{2} n^{2 \tau}\right\} &~\leq \mathbb{P}\left\{I_{1} \geq \frac{c_{1}^{2} n^{2 \tau}}{2}\right\}+\mathbb{P}\left\{I_{3} \geq \frac{c_{1}^{2} n^{2 \tau}}{2}, \mathcal{M} \text { holds }\right\}+\mathbb{P}\left(\mathcal{M}^{c}\right)\\
&~\leq \exp\left(-C_2n^{B(\gamma,r,\eta,\breve{\eta})}\right),
\end{align*}
where
{
\begin{align*}
&B(\gamma,r,\eta,\breve{\eta}) \\
&\leq\left\{\begin{array}{l}
{\gamma (\breve{\eta}-3r/2)\wedge(1-2\eta)\wedge(1-r+\eta)\gamma/2},\\
{ \qquad\text { if } \gamma < 2 \text { and }  r\geq \delta_2\vee 2\tau-\delta_3 \text{and }{\eta}>\delta_1}\\
{\gamma (\breve{\eta}-3r/2)\wedge(1-2r)\gamma/4\wedge(1-2{\eta}), }\\
{\qquad\text { if } \gamma < 2 \text { and }  2\tau-\delta_3<r< \delta_2 \text{ and } {\eta}>\delta_1} \\
{\gamma (\breve{\eta}-3r/2)\wedge(1-2\eta)\wedge(2\tau-r+1)\gamma/4,}\\
{ \qquad\text { if } \gamma < 2 \text { and }  \delta_2<r\leq 2\tau-\delta_3 \text{ and } \eta>\delta_1} \\
{\gamma (\breve{\eta}-3r/2)\wedge(1-2r)\gamma/4\wedge(1-2\eta)\wedge(2\tau-r+1)\gamma/4, }\\
{\qquad\text { if } \gamma < 2 \text { and } r< \delta_2\wedge2\tau-\delta_3  \text{ and } \eta>\delta_1} \\
{\gamma (\breve{\eta}-3r/2)\wedge(1-\eta)\gamma/2\wedge(2\tau-r+1)\gamma/4, }\\
{\qquad \text { if } \gamma < 2 \text { and } \delta_2<r< 2\tau-\delta_3 \text{ and } \eta<\delta_1}\\
{\gamma (\breve{\eta}-3r/2)\wedge(1-\eta)\gamma/2\wedge(1-2r)\gamma/4\wedge(2\tau-r+1)\gamma/4,} \\
{\qquad\text { if } \gamma < 2 \text { and } r< \delta_2\wedge2\tau-\delta_3  \text{ and } \eta<\delta_1}\\
{\gamma (\breve{\eta}-3r/2)\wedge(1-2{\eta}),}
{\text { if } 2\leq\gamma <4 \text { and } r\geq \delta_2}\\
{\gamma (\breve{\eta}-3r/2)\wedge(1-2r)\gamma/4\wedge(1-2{\eta}), }
{\text { if } 2\leq\gamma <4 \text { and } r<\delta_2}\\
{\gamma (\breve{\eta}-3r/2)\wedge(1-2{\eta}),}
{ \text { if } \gamma\geq 4}\end{array}\right.
\end{align*}
}
with $\gamma=\frac{\gamma_1\gamma_2}{\gamma_1+\gamma_2}$, $\breve{\eta}=\eta+\frac{2\tau}{3}-\frac{1}{3}$, $\delta_1=\frac{1}{2}-\frac{\gamma}{8-2\gamma}$ and $\delta_2=\frac{1}{4}-\frac{\gamma}{32-4\gamma}$ and $\delta_3=1-2\delta_1$.
Then we can directly get the result in the Theorem.
\qed

\

\textbf{Proof of Lemma~4.2.}
Note that  $ l^c_j(\bm 0)=2\sum_{i=1}^n\log\{1+\bm\alpha^\top g^c_{ij}(\bm 0)\}$, where $\bm\alpha$ is the Lagrange multiplier satisfying
$ \bm0 = \sum_{i=1}^n\frac{g^c_{ij}(\bm 0)}{1+\bm\alpha^\top g^c_{ij}(\bm 0)}$ and $g^c_{ij}(\bm 0)=[X_{ij} - \mathbb{E}(X_j|\BC^\top\mathbf{X}_{i\mathcal{C}})]\mathbf{y}_i$.
 According to Condition (C1') and Theorem~4.1, we can easily prove the result.
\qed

\

\textbf{Proof of Theorem~4.3.}
To simplify the notation, we use $\hat U_{i,jk}=[X_{ij}-\widehat {\mathbb{E}}(X_j|{\bf x}_{i\mathcal{C}})]Y_{ik}$. Similar to the proof of Lemma 4 in \cite{hu2017conditional} and using the similar technique in the proof of Lemma~4.1, we can prove that 
\begin{align*}
&\mathbb{P}\{\widehat{l_j^c}(\bm 0)<2 t\} \leq \mathbb{P}\left\{\sum_{i=1}^{n}\sum_{k=1}^q \hat U_{i, jk}<(t n^{\epsilon}+n^{1-\epsilon}) \max _{i,k}\left|\hat U_{i,jk}\right|\right\}\\
&\leq\mathbb{P}\left\{\sum_{i=1}^{n}\sum_{k=1}^q  U_{i, jk}<(t n^{\epsilon}+n^{1-\epsilon}) \max _{i,k}\left|\hat U_{i,jk}\right|+nq\max_{i,k}|\hat U_{i, jk}-U_{i, jk}|\right\}\\
&\leq\mathbb{P}\left\{\sum_{i=1}^{n}\sum_{k=1}^q  U_{i, jk}<(t n^{\epsilon}+n^{1-\epsilon}) \max _{i,k}\left|U_{i,jk}\right|+(t n^{\epsilon}+n^{1-\epsilon}+nq)\max_{i,k}|\hat U_{i, jk}-U_{i, jk}|\right\}\\
&\leq \mathbb{P}\left\{\sum_{i=1}^{n}\sum_{k=1}^q U_{i, jk}-n\sum_{k=1}^q \mu_{jk}<(t n^{\epsilon}+n^{1-\epsilon})M -n\sum_{k=1}^q \mu_{jk}+(t n^{\epsilon}+n^{1-\epsilon}+nq)\max_{i,k}|\hat U_{i, jk}-U_{i, jk}|\right\}\\
&\quad+\mathbb{P}\left\{ \max _{i,k}\left|U_{i,jk}\right|>M\right\}\\
&\leq \sum_{k=1}^q \mathbb{P}\left\{\sum_{i=1}^{n} (U_{i, jk}-\mu_{jk})<\frac{(t n^{\epsilon}+n^{1-\epsilon})M-n\sum_{k=1}^q \mu_{jk}+(t n^{\epsilon}+n^{1-\epsilon}+nq)\max_{i,k}|\hat U_{i, jk}-U_{i, jk}|}{q} \right\}\\
&\quad+\mathbb{P}\left\{ \max _{i,k}\left|U_{i,jk}\right|>M\right\}\\
&=\sum_{k=1}^q \mathbb{P}\{\frac{1}{n^\frac{1}{2}\sigma_{jk}}\sum_{i=1}^{n} (U_{i, jk}-\mu_{jk})<\frac{(t n^{\epsilon-\frac{1}{2}}+n^{\frac{1}{2}-\epsilon})M-n^\frac{1}{2}\sum_{k=1}^q \mu_{jk}}{q\sigma_{jk}}\\
&\quad + \frac{(t n^{\epsilon}+n^{1-\epsilon}+nq)\max_{i,k}|\hat U_{i, jk}-U_{i, jk}|}{n^{\frac{1}{2}}q\sigma_{jk}}\}+\mathbb{P}\left\{ \max _{i,k}\left|U_{i,jk}\right|>M\right\},
\end{align*}
where $U_{i,jk}=[X_{ij}-{\mathbb{E}}(X_j|{\mathcal{B}}_\C^\top\X_{i\mathcal{C}})]Y_{ik}$ and  $\mu_{jk}=\E(U_{i,jk})$.
{Since $\widehat{\mathcal{B}}_\C$ is the estimate of $\BC$ by SIR, and $\widehat{\mathbb{E}}(X_j|\widehat{\mathcal{B}}_\C^\top\X_{i\C})$ is estimated as
	$
	\X_{i\C}^\top\widehat{\mathcal{B}}_\C \widehat{\mbox{cov}}
	\left(\widehat{\mathcal{B}}_\C^\top\X_{i\C}\right)^{-1}\widehat{\mbox{cov}}(\widehat{\mathcal{B}}_\C^\top\X_{i\C}, X_j).
	$
} Following the similar analysis of the proof of lemma 4 in \cite{hu2017conditional}, we can get $\max_{i,k}|\hat U_{i, jk}-U_{i, jk}|=O_p(n^{\omega-\frac{1}{2}})$, and then
$$\frac{(t n^{\epsilon}+n^{1-\epsilon}+nq)\max_{i,k}|\hat U_{i, jk}-U_{i, jk}|}{n^{\frac{1}{2}}q\sigma_{jk}}=O_p(n^{\frac{1}{2}}\max_{i,k}|\hat U_{i, jk}-U_{i, jk}|)=O_p(n^{\omega}),$$
and
$$\frac{(t n^{\epsilon-\frac{1}{2}}+n^{\frac{1}{2}-\epsilon})M-n^\frac{1}{2}\sum_{k=1}^q \mu_{jk}}{q\sigma_{jk}}=O_p(n^{\frac{1}{2}}\max_{k}|\mu_{jk}|).$$
Moreover, under Condition (C1'), we can get $n^{\frac{1}{2}}|\mu_{jk}|\geq c_1n^{\frac{1}{2}-\kappa}$ for any $j\in \DA$ and $1\leq k\leq q$, hence $n^\omega=o_p(n^{\frac{1}{2}}\max_{k}|\mu_{jk}|)$ following by our assumption $\kappa\leq \frac{1}{2}-\omega$. This implies that we can neglect the item $$\frac{(t n^{\epsilon}+n^{1-\epsilon}+nq)\max_{i,k}|\hat U_{i, jk}-U_{i, jk}|}{n^{\frac{1}{2}}q\sigma_{jk}}$$or replace it by $O_p(n^{\frac{1}{2}}\max_{k}|\mu_{jk}|)$. Hence, it has that
\begin{align*}
\mathbb{P}\{\widehat{l_j^c}(\bm 0)<2 t\}\leq &\sum_{k=1}^q \mathbb{P}\left\{\frac{1}{n^\frac{1}{2}\sigma_{jk}}\sum_{i=1}^{n} (U_{i, jk}-\mu_{jk})<\frac{(t n^{\epsilon-\frac{1}{2}}+n^{\frac{1}{2}-\epsilon})M-n^\frac{1}{2}\sum_{k=1}^q \mu_{jk}}{q\sigma_{jk}}\right\}\\
&+\mathbb{P}\left\{ \max _{i,k}\left|U_{i,jk}\right|>M\right\},
\end{align*}
Therefore, we can get the results based on the proof of Lemma~4.1.
\qed

\textbf{Proof of Theorem~4.4.}

    {First, according to the proof of Theorem~4.2, we can prove that}
{
 \begin{align*}
 &\mathbb{P}\left\{{l_j^c}(\bm 0) \geq c_{3}^{2} n^{2 \tau}\right\}\\
\leq~ & \mathbb{P}\left\{{I_{1}} \geq \frac{c_{3}^{2} n^{2 \tau}}{2}\right\}+\mathbb{P}\left\{{I_{3}} \geq \frac{c_{3}^{2} n^{2 \tau}}{2}, { \mathcal{M}} \text { holds }\right\}+\mathbb{P}\left\{{\mathcal{M}}^{c}\right\}\\
\leq~  & q\mathbb{P}\left\{\left|\frac{1}{n}\sum_{i=1}^n U_{i, jk}\right|\geq   \sqrt{\frac{\lambda_{\min}}{2}}c_3n^{\tau-\frac{r}{2}-\frac12}\right\}+\mathbb{P}\left(\lambda_{\min}(\bm S_j)<\lambda_{\min}\right)\\
  &+2\mathbb{P}\left\{\bm\theta^\top\bm  S_j\bm\theta<\lambda_{\min}\right\}+\mathbb{P}\left\{\max_{l}\|
   \bm u_{lj}\|_2\geq n^{\breve{e}}\right\}+\mathbb{P}\left\{\max_{l}\|
   \bm u_{lj}\|_2\geq \frac14 n^{e}\right\}\\
  &+\mathbb{P}\left\{\bm\theta^\top \bar{ {\bm u}}_{j} \geq \tilde c_{1}n^{2 \tau/3-1/3-\breve{e}}\right\}+\mathbb{P}\left\{\bm\theta^\top \bar{ {\bm u}}_{j}\geq n^{-e}\lambda_{\min}\right\}\\
=: ~  &\R_1+\R_2+\R_3+\R_{41}+\R_{42}+\R_{51}+\R_{52},
\end{align*}
    where $e\in (0,\eta-r)$,  $\breve{e}\in (0,\eta+\frac{2}{3}\tau-\frac{1}{3}-r)$. }

{
Note that  $\widehat{l_j^c}(\bm 0)=2\sum_{i=1}^n\log\{1+ \hat {\bm \alpha}^\top\hat{\bm u}_{ij}\}$, where  $\hat {\bm \alpha}$ satisfies $\bm 0=\sum_{i=1}^{n} \frac{\hat {\bm u}_{i j}}{1+\bm\alpha^\top  \hat {\bm u}_{i j}}$, $\hat {\bm {u}}_{i j}=(\hat U_{i,j1},\ldots,\hat U_{i,jq})^\top $and $\hat U_{i,jk}=[X_{ij}-{\widehat {\mathbb{E}}}(X_j|{\widehat {\mathcal{B}}}_\C^\top\X_{i\mathcal{C}})]Y_{ik}$.  With the similar technique in the proof of Theorem~4.2, we can get the sample version as following
\begin{align*}
 &\mathbb{P}\left\{\widehat{l_j^c}(\bm 0) \geq c_{3}^{2} n^{2 \tau}\right\}\\
\leq~ & \mathbb{P}\left\{\hat {I_{1}} \geq \frac{c_{3}^{2} n^{2 \tau}}{2}\right\}+\mathbb{P}\left\{\hat {I_{3}} \geq \frac{c_{3}^{2} n^{2 \tau}}{2}, \widehat{ \mathcal{M}} \text { holds }\right\}+\mathbb{P}\left\{\widehat{\mathcal{M}}^{c}\right\}\\
  \leq ~& q\mathbb{P}\left\{\left|\frac{1}{n}\sum_{i=1}^n \hat U_{i, jk}\right|\geq  \sqrt{\frac{\lambda_{\min}}{2}}c_3n^{\tau-\frac{r}{2}-\frac12}\right\}+\mathbb{P}\left(\lambda_{\min}(\widehat{\bm S}_j)<\lambda_{\min}\right)\\
  &+2\mathbb{P}\left\{\bm\theta^\top\widehat{\bm S}_j\bm\theta<\lambda_{\min}\right\}+\mathbb{P}\left\{\max_{l}\|
   \hat{\bm u}_{lj}\|_2\geq n^{\breve{e}}\right\}+\mathbb{P}\left\{\max_{l}\|
   \hat{\bm u}_{lj}\|_2\geq \frac14 n^{e}\right\}\\
  &+\mathbb{P}\left\{\bm\theta^\top \hat{\bar{\bm u}}_{j} \geq \tilde c_{1}n^{2 \tau/3-1/3-\breve{e}}\right\}+\mathbb{P}\left\{\bm\theta^\top \hat{\bar{ \bm u}}_{j}\geq n^{-e}\lambda_{\min}\right\}\\
     =:~&\hR_1+\hR_2+\hR_3+\hR_{41}+\hR_{42}+\hR_{51}+\hR_{52},
\end{align*}
where $\hat I_1$, $\hat I_3$, $\hM$, $\bar{\hat{ \bm u}}_{j}$ and $\hat{\bm S}_j$ are the sample version of $I_1$, $I_3$, $\M$, $\bar{{ \bm u}}_{j}$ and $\bm S_j$, respectively.}

{
In fact, $\mathbb{P}\left\{\widehat{l_j^c}(\bm 0)\geq c_{3}^{2} n^{2 \tau}\right\}$ has the same upper bound as $\mathbb{P}\left\{{l_j^c}(\bm 0)\geq c_{3}^{2} n^{2 \tau}\right\}$.
Since 
$$\left|\frac1n\sum_{i=1}^n\hat U_{i,jk}\right|\leq \left|\frac1n\sum_{i=1}^n U_{i,jk}\right|+\max_{i,k}\left|\hat U_{i,jk}-U_{i,jk}\right|$$
and
$$\max_{i,k}\left|\hat U_{i,jk}-U_{i,jk}\right|=O_p(n^{\omega-1/2}),$$
then
\begin{align*}
\hR_1&~=q\mathbb{P}\left(\left|\frac1n\sum_{i=1}^n\hat U_{i,jk}\right|\geq \sqrt{\frac{\lambda_{\min}}{2}}c_3n^{\tau-\frac{r}{2}-\frac12}\right)\\
&~\leq q\mathbb{P}\left(\left|\frac1n\sum_{i=1}^n\hat U_{i,jk}\right|+\max_{i,k}\left|\hat U_{i,jk}-U_{i,jk}\right|\geq \sqrt{\frac{\lambda_{\min}}{2}}c_3n^{\tau-\frac{r}{2}-\frac12}\right)\\
&~= q\mathbb{P}\left(\left|\frac1n\sum_{i=1}^n\hat U_{i,jk}\right|\geq \sqrt{\frac{\lambda_{\min}}{2}}c_3n^{\tau-\frac{r}{2}-\frac12}-\max_{i,k}\left|\hat U_{i,jk}-U_{i,jk}\right|\right)
\end{align*}
Then, we can choose $\widehat{C}_3$ such that $$\sqrt{\frac{\lambda_{\min}}{2}}c_3n^{\tau-\frac{r}{2}-\frac12}-\max_{i,k}\left|\hat U_{i,jk}-U_{i,jk}\right|\geq \widehat{C}_3n^{\tau-\frac{r}{2}-\frac12},$$ when $\tau-\frac{r}{2}-\frac12\geq \omega-\frac{1}{2}.$
As a result
\begin{equation*}
	\hR_1\leq q\mathbb{P}\left(\left|\frac1n\sum_{i=1}^n\hat U_{i,jk}\right|\geq \widehat{C}_3n^{\tau-\frac{r}{2}-\frac12} \right).
\end{equation*}
    According to the proof of Theorem 4.2 and neglecting the coefficient,  we can conclude that $\hR_1$ and $\R_1$ have the same upper bound with the same order. }

{
Due to the same technique, we can get that $\hR_{51}$ and $\R_{51}$, $\hR_{52}$ and $\R_{52}$ have the same upper bound with the same order respectively when  $\eta<\frac{1}{2}-\omega$.
On the other hand,  $\|\hat {\boldsymbol u}_{lj}\|_2\leq  \|{\boldsymbol u}_{lj}\|_2+\|\hat {\boldsymbol u}_{lj}-{\boldsymbol u}_{lj}\|_2$ and $\|\hat {\boldsymbol u}_{lj}-{\boldsymbol u}_{lj}\|_2\leq q^{1/2}\max_{l,k}|\hat U_{l, jk}- U_{l, jk}|$.  Since $r<\frac{2}{7}\eta$ and $\eta<\frac{1}{2}-\omega$, then we can get $\frac{r}{2}+\omega-\frac{1}{2}<0$, and  $\hR_{41}$ and $\R_{41}$  and $\hR_{42}$ and $\R_{42}$ have the same upper bound with the same order, respectively. 
  In addition, $$\theta^T \bm S_j\theta+\lambda_{\min}(\triangle \bm S_j)\leq\theta^T\hat S_j\theta=\theta^T\bm S_j\theta+\theta^T \triangle \bm S_j\theta\leq \theta^T\bm S_j\theta+\lambda_{\max}(\triangle \bm S_j) $$
where $\theta$ is an unit vector, $\lambda_{\min}(\triangle \bm S_j),\lambda_{\max}(\triangle S_j)$ denote the maximum and minimum eigenvalue of matrix $\triangle S_j$ and $(\triangle S_j)_{kt}=(\hat S_j-S_j)_{kt}=\frac{1}{n}\sum_{i=1}^n(\hat U_{ij,k}\hat U_{ij,t}-U_{ij,k}U_{ij,t})$ and $k,t=1,\ldots,q.$  Since, 
$$\max {|\lambda(\triangle \bm S_j)|} \leq \max_{k}{\sum^q_{t=1}(\triangle \bm S_j)_{kt}}\leq q\max_{k,t} |\hat U_{ij,k}\hat U_{ij,t}-U_{ij,k}U_{ij,t}|=O_p(n^{r+2\omega-1/2})$$
where $\lambda(\triangle \bm S_j)$ denote the eigenvalue of matrix $\triangle \bm S_j$. Hence, we can get 
\begin{align*}
\hR_3&~=\mathbb{P}\left\{\bm\theta^\top\widehat{\bm S}_j\bm\theta<\lambda_{\min}\right\}\\
&\leq\mathbb{P}\left\{\bm\theta^\top{\bm S}_j\bm\theta+\lambda_{\min} ({\triangle \bm S_j})<\lambda_{\min}\right\}\\
&=\mathbb{P}\left\{\bm\theta^\top{\bm S}_j\bm\theta<\lambda_{\min}-\lambda_{\min} ({\triangle \bm S_j})\right\}\\
&\leq\left\{\begin{array}{l}
\mathbb{P}\left\{\bm\theta^\top{\bm S}_j\bm\theta<\lambda_{\min}\right\}\\
\qquad\text { if } \lambda_{\min} ({\triangle \bm S_j})>0, \\
\mathbb{P}\left\{\bm\theta^\top{\bm S}_j\bm\theta<\lambda_{\min}+O_p(n^{r+2\omega-1/2})\right\}\\
 \qquad\text { if } \lambda_{\min} ({\triangle \bm S_j})<0 .\end{array}\right.
\end{align*}
Therefore, when $r+2\omega-1/2<0$, we can obtain $\hR_{3}$ and $\R_{3}$ have the same upper bound with the same order. And through the similar technique, we can get that the same result for  $\hR_{2}$ and $\R_{2}$. And then following by the similar argument in the first part of proof of the Theorem~4.2, 
we can  directly obtain the result:
$$\mathbb{P}\left\{|\widehat{\DA}_{\gamma_{n}}|>s_{\DA}\right\}\leq p_1\exp\left(-C_5n^{B(\gamma,r,\eta,\breve{\eta})}\right),$$
where $p_1$ is the size of $X_\D$, 
and $\omega$ satisfies $\mi |X_{ij}Y_{ik}|=O_p(n^\omega)$ and $\omega<\frac12-\eta$, for $j\in\C$	and  $r$ satisfies the dimension $q$ of response satisfies $q(n)=O(n^r)$ and $0\leq r<\frac{2}{7}\eta\wedge\frac{1}{2}-\kappa\wedge(\frac{1}{2}-2\omega)$, 
    	 $\gamma$, $\delta_1$, $\delta_2$ and $\delta_3$ are given in Theorem 4.2, $C_5$ depends only on $K_1, K_2,\gamma_1$ and $\gamma_2$ given in Condition (C2).}
	
\qed

\newpage

	\begin{table}[!h]
		\centering
		\caption{The quartiles of minimum model size of the selected models in Example 4.2.}\label{ms_ex2}
		\scalebox{0.95}{
		\begin{tabular}{lrrrrrrrrrr}
			\toprule
			&\multicolumn{5}{c}{$\sigma_i(\mathbf{X})$:~case (a)}&\multicolumn{5}{c}{$\sigma_i(\mathbf{X})$:~case (b)}\\
			\cmidrule(lr){2-6}\cmidrule(lr){7-11}
			Method &$5\%$&$25\%$&$50\%$&$75\%$&$95\%$&  $5\%$&$25\%$&$50\%$&$75\%$&$95\%$ \\  \midrule
			\multicolumn{11}{l}{$\rho=0$}\\
			DCSIS        & 5.0  &10.0   &35.5   &116.0  &381.1 &487.9 &756.5 &898.0  &966.3 &997.0\\
			BCorSIS      &5.0 &5.0 &6.0&9.0& 34.1&38.7&109.0&229.5&376.8& 671.8\\
			PS           & 5.0  &5.0    &5.0    &5.0    &5.0     &22.0   &158.8 &360.0   &660.5  &936.5\\
			RCC$_{sp}$   & 5.0  &5.0    &5.0    &5.0    &11      &6.0 & 15.0 & 32.8 & 58.2 & 209.1 \\
            RCC$_{kd}$   & 5.0  &5.0    &5.0    &5.0    &6.2     &6.0 & 14.0 & 27.6 & 61.2 & 168.3 \\
            ELSIS$_{avg}$& 5.0  &5.0    &7.0    &20.3   &94.1    &6.0   &12.0  &36.0  &113.0  &523.2\\
			ELSIS$_{max}$& 6.0  &14.0   &41.0   &129.0  &423.4   &6.0   &10.8  &26.0  &51.3   &159.1\\
			MELSIS       & 5.0  &5.0    &5.0    &5.0    &8.0     &6.0   &11.0  &22.0  &52.0   &131.0\\
			\multicolumn{11}{l}{$\rho=0.5$}\\
			DCSIS        & 5.0  &9.0    &27.0   &98.0   &317.6 &546.7& 763.8 & 887.5 & 957.0 & 998.1 \\
			BCorSIS      & 5.0  & 5.0   &6.0    &11.0   &48.2  &16.9 & 77.2& 186.5& 299.3& 521.2  \\
			PS           & 5.0  &5.0    &5.0    &5.0    &5.0   &23.0 & 157.3 & 357.5 & 592.3 & 896.4 \\
			RCC$_{sp}$   & 5.0  &5.0    &5.0    &5.0    &7.1   &5.0  &28.0&36.5&55.5&231.4  \\
            RCC$_{kd}$   & 5.0  &5.0    &5.0    &5.0    &7.0   &5.0  &27.0&35.0&47.2&236.5  \\
			ELSIS$_{avg}$& 5.0  &5.0    &7.0    &15.0   &74.1  &7.0  & 19.8 & 47.0 & 225.3 & 627.2 \\
			ELSIS$_{max}$& 6.0  &15.0   &52.0   &141.3  &466.3 &5.0  &11.0 &26.0 &60.3 &192.2\\
			MELSIS       & 5.0  &5.0    &5.0    &5.0    &6.0   &5.0  &20.0 &24.0 &63.0 &175.4\\
			\bottomrule
		\end{tabular}
		}
	\end{table}

\begin{table}[!h]
	\centering
	\caption{The quartiles of minimum model size of the selected model in Example 4.3.}\label{ms_ex3}
	\scalebox{0.95}{
	\begin{tabular}{lcrrrrrrrrrr}
		\toprule
		&&\multicolumn{5}{c}{$\sigma_i(\mathbf{X})$:~case (a)}&\multicolumn{5}{c}{$\sigma_i(\mathbf{X})$:~case (b)}\\
		\cmidrule(lr){3-7}\cmidrule(lr){8-12}
		Method & &$5\%$&$25\%$&$50\%$&$75\%$&$95\%$&$5\%$&$25\%$&$50\%$&$75\%$&$95\%$ \\  \midrule
		DCSIS&&492.2&906.5&988.0&999.0&1000.0&357.1&776.0&  971.5& 999.0&1000.0\\
		BCorSIS&&286.3&695.0& 918.0& 982.5& 1000.0&320.2& 695.0&888.0& 985.3& 1000.0 \\
        RCC$_{sp}$&&69.0&484.3&764.5&941.3&995.2&124.8&467.5&757.0&940.3&998.0 \\
        RCC$_{kd}$&&152.8&580.3&848.0&980.0&1000&168.8&618.3&873.0&975.5&1000.0 \\
		MELSIS&&196.7&398.8&667.5&867.5&992.05&196.9&595.5&882.5&988.5&1000.0\\
		CMELSIS
		&$\mathcal{C}_1$&2.0&2.0&2.0&2.0&2.0&2.0&2.0&2.0&2.0&6.0\\
		&$\mathcal{C}_2$&2.0&2.0&2.0&2.0&2.0&2.0&2.0&2.0&2.0&4.0\\
		&$\mathcal{C}_3$&3.0&3.0&3.0&3.0&6.0&3.0&3.0&3.0&5.0&33.2\\
		&$\mathcal{C}_4$&4.0&4.0 &4.0  &4.0 &8.4 &4.0&4.0&4.0&10.0&94.4 \\
		CELSIS$_{avg}$
		&$\mathcal{C}_1$&2.0&2.0 &2.0  &10.0 &39.1 &2.0 &2.0  &3.0  &10.0 &52.3  \\
		&$\mathcal{C}_2$&8.9&59.5&187.5&397.5&660.9&38.9&116.8&256.0&455.0&773.6 \\
		&$\mathcal{C}_3$&3.0&10.0&34.0 &65.8 &192.1&3.0 &14.0 &44.5 &104.8&308.5 \\
		&$\mathcal{C}_4$&5.0&14.0&40.0 &103.0&268.3&4.0 &17.5 &88.0 &250.5&712.1 \\
		CELSIS$_{max}$
		&$\mathcal{C}_1$&2.0&2.0 &2.0  &3.0  &9.2  &2.0 &2.0  &2.0  &3.0  &18.0  \\
		&$\mathcal{C}_2$&2.0&5.0 &26.5 &129.0&485.9&2.0 &12.8 &52.0 &169.8&655.4 \\
		&$\mathcal{C}_3$&3.0&3.0 &8.5  &28.3 &170.1&3.0 &5.0  &12.5 &50.8 &251.9 \\
		&$\mathcal{C}_4$&4.0&5.0 &9.0  &32.25&127.0&4.0 &9.8  &63.5 &205.5&783.8 \\		\bottomrule
	\end{tabular}
	}
\end{table}

\begin{table}[!h]
\centering
\caption{Simulation results of Example 4.4.}\label{two-stage}
\begin{tabular*}{\hsize}{@{}@{\extracolsep{\fill}}lccccc@{}}
\toprule
Method & No. of variables in conditional set &$d=21$ & $d=31$ & $d=42$ \\ \midrule
MELSIS-CMELSIS  				& 3 & 0.99 & 1.00 & 1.00 \\
                            & 5 & 0.99 & 1.00 & 1.00 \\
                           	& 7 & 1.00 & 1.00 & 1.00 \\
                           	& 9 & 1.00 & 1.00 & 1.00 \\
ELSIS$_{avg}$-CELSIS$_{avg}$& 3 & 0.41 & 0.47 & 0.49 \\
							& 5 & 0.54 & 0.56 & 0.59 \\
							& 7 & 0.58 & 0.61 & 0.62 \\
							& 9 & 0.61 & 0.61 & 0.61 \\
ELSIS$_{max}$-CELSIS$_{max}$& 3 & 0.51 & 0.53 & 0.55 \\
							& 5 & 0.65 & 0.69 & 0.70 \\
							& 7 & 0.66 & 0.69 & 0.70 \\
							& 9 & 0.68 & 0.71 & 0.75 \\	
\bottomrule
\end{tabular*}
\end{table}

\begin{table}[!h]
\centering
\caption{The selected SNPs by MELSIS, ELSIS$_{avg}$, ELSIS$_{max}$ and PS}\label{snps}
\scalebox{0.9}{
\begin{tabular}{llllll}
\toprule
Method & The survived SNPs \\ \midrule
MELSIS
& D14Mit3  & D14Cebrp312s2 & D14Rat52 & D14Mit8  & D4Rat7 \\
& D3Mit6   & D4Rat252      & D14Mit9  & Es13     & D10Rat226 \\
& D6Rat132 & D1Rat327      & D14Utr6  & D14Rat77 & D4Rat10   \\
& D8Rat135 & D14Rat36      & D14Utr7  & D4Rat151 & D6Cebrp97s14\\
RCC$_{sp}$
& D14Mit3 & D14Cebrp312s2 & D14Rat52 & D14Mit8 & D14Mit9 \\
& D16Rat72 & D3Mit6 & D14Utr6 & D14Rat77 & D14Rat36\\
& D14Utr2 & D14Utr8 & D4Rat152 & D1Rat327 & D14Utr7 \\
& D4Rat102 & D16Mit5 & D14Rat90 & Pthlh & D2Rat69 \\
RCC$_{kd}$
& D14Mit3 & D14Cebrp312s2 & D14Rat52 & D4Rat152 & D14Mit8 \\
& D14Mit9 & D14Rat36 & Lep & D14Utr2 & D4Rat102 \\
& D14Utr8 & Pthlh & D14Utr6 & D14Rat77 & D16Rat72 \\
& D4Rat16 & D4Utr1 & D3Mit6 & D4Mit9 & D16Mit5 \\
ELSIS$_{avg}$
&Prl           &Rbp2        &D17Mit3    &D17Rat17       &D1Rat270\\
&D1Cebr103s1   &D20Rat55    &D8Mgh4     &D8Cebr204s21   &D1Rat277\\
&D11Rat47      &D20Mit1     &D1Rat42    &D8Cebr46s2     &D15Utr2 \\
&D15Rat21      &D20Rat19    &D8Rat135   &D3Mit17        &D20Rat52  \\
ELSIS$_{max}$
&Prl           &Rbp2        &D17Mit3       &D17Rat17       &D1Rat270    \\
&D1Cebr103s1   &D1Rat277    &D11Rat47      &D8Cebr204s21   &D8Cebr46s2  \\
&D8Mgh4        &D20Rat55    &D20Mit1       &D15Utr2        &D1Rat42     \\
&D1Rat55       &D1Arb17     &D1Cebr21s2    &D3Cebr4s5      &D8Rat135    \\
&D20Rat52      &Inha        &D10Mit3       &D20Rat19       &D15Rat21    \\
&D3Mit17       &D15Rat123   &D15Cebr7s13   &D15Rat68      \\
PS
&D14Mit3    &D14Cebrp312s2   &D14Rat52   &D14Mit9        &D14Rat36     \\
&D14Mit8    &D16Rat72        &D14Utr2    &D3Mit6         &D16Mit5      \\
&D14Utr6    &D14Rat77        &D4Rat252   &D14Utr7        &D4Rat152     \\
&D14Utr8    &D14Rat90        &D11Mit4    &D4Rat102       &D6Rat132     \\
\bottomrule
\end{tabular}}
\end{table}

\end{document}